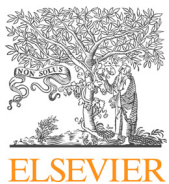
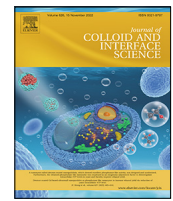

# Hydrodynamic interactions between charged and uncharged Brownian colloids at a fluid-fluid interface

Archit Dani [c,*], Mohsen Yeganeh [b], Charles Maldarelli [a]

[a] *Levich Institute and Department of Chemical Engineering, City College of the City University of New York, New York, NY 10031, USA*
[b] *ExxonMobil Research and Engineering Company, Annandale, NJ, USA*
[c] *Intel Corporation, Hillsboro, OR 97123, USA*

## HIGHLIGHTS

- Hydrodynamic interactions (*HI*) affect the aggregation dynamics of colloids straddling a fluid–fluid interface.
- For uncharged particles, *HI* decrease the probability of aggregate formation for all Péclet (*Pe*) numbers.
- For charged particles, *HI* change the probability of aggregate formation depending on the relative magnitude and influence of electrostatic repulsive interactions.

## GRAPHICAL ABSTRACT

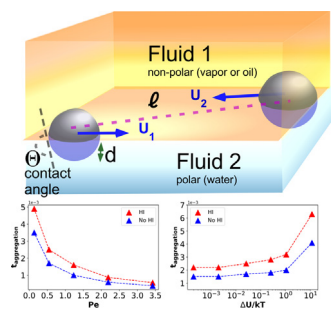

## ABSTRACT

*Hypothesis:* The cluster formation and self-assembly of floating colloids at a fluid/fluid interface is a delicate force balance involving deterministic lateral interaction forces, viscous resistance to relative colloid motion along the surface and thermal (Brownian) fluctuations. As the colloid dimensions get smaller, thermal forces and associated drag forces become important and can affect the self assembly into ordered patterns and crystal structures that are the starting point for various materials applications.
*Numerics:* Langevin dynamic simulations for particle pairs straddling a liquid–liquid interface with a high viscosity contrast are presented to describe the lateral interfacial assembly of particles in Brownian and non-Brownian dominated regimes. These simulations incorporate capillary attraction, electrostatic repulsion, thermal fluctuations and hydrodynamic interactions (*HI*) between particles (including the effect of the particle immersion depth). Simulation results are presented for neutrally wetted particles which form a contact angle $\theta = 90^0$ at the interface.
*Findings:* The simulation results suggest that clustering, fractal growth and particle ordering become favorable outcomes at critically large values of the *Pe* numbers, while smaller *Pe* numbers exhibit higher probabilities of final configurations where particle motion remains uncorrelated in space and particle pairs are found to be more widely separated especially upon the introduction of *HI*.

© 2022 Elsevier Inc. All rights reserved.

## 1. Introduction

Colloids have the ability to separate out from immiscible fluid phases bounding an interface and adsorb onto the interface forming a stable monolayer (for reviews cf. [1–6]). The physics of colloids at these interfaces is markedly different from colloids in the bulk. For a (spherical) colloid situated at the interface separating a nonpolar phase (e.g. air or oil) from a polar phase (e.g. water), the interfacial energy of adsorption $E_{ads}$ is easily calculated to be $\Delta E_{ads} = -\pi a^2 \gamma (1 \pm \cos\theta)^2$, where $a$ is the colloid radius, $\gamma$ is the interfacial tension, and $\theta$ is the three phase contact angle (measured in the polar phase). The "+(-)" sign is the energy relative to adsorption from the nonpolar (polar) phase. $\theta$ is related to the equilibrium immersion depth $d$ as given by $\frac{d}{a} = 1 + \cos\theta$, where $d$ is measured from the interface to the bottom of the colloid. When the surface of the colloid is partially wetting to both phases ($|\cos(\theta)| \neq 1$), the adsorption energy is greater than thermal energy, thereby favoring irreversible adsorption at fluid/fluid interfaces. When these particles adsorb on a fluid–fluid interface forming a stable monolayer, particle detachment is possible only by application of a large external force [7–10].

Recent attention has focused on the self-organization of the particles at the interface, and the structure of the resulting monolayer assemblies in different areas of research such as pollination [11], locomotion of microorganisms [12], froth flotation processes in the mining industry and in the bottom up assembly of materials from nanoscale [13,14] to mesoscale objects [15]. Experimental studies at fluid–fluid interfaces have shown how ordered structures can be formed and further be utilized in applications such as coatings (superhydrophobic or antireflection), templates for nanosphere lithography [16] and for optoelectronic devices [17]. Such studies on interfacial self-assembly of colloids depends on a complex interplay of different parameters such as compression rate, wettability, particle charge, size [18–20], electrolyte [21–27] and interfacial curvature [28,29].

The self-assembled interfacial configurations of colloids are driven by deterministic attractive interactions which act between the particles and along the surface. These attractive interactions can be of multiple types such as van der Waals attractions [30], gravity-driven capillary attractive interactions [31–41] and multipole attractive interactions stemming from asperities on the colloid surface [42,43,40,44–46] lead to undulating contact lines, overlapping interface deformations and capillary attraction. In addition, if the particles are charged, strong dipolar electrostatic repulsions between the particles are present [47–49], and the action of the charges also causes the particles to be pushed into the phase of higher dielectric constant, resulting in a capillary attraction (electrodipping force) [50,48,51–56,33]. Similarly, colloids with magnetic domains [57], which are forced to align in a magnetic field applied perpendicular to the surface, exhibit repulsive lateral interactions. Structures can also be assembled by the compression of colloidal monolayers adsorbed at a surface, in which case the attractive and repulsive deterministic forces as well as the surface compression force balance to determine the structure. For either self- or imposed assembly, colloidal structures can only form if the deterministic lateral forces are much larger than the stochastic Brownian or thermal fluctuation forces.

The colloidal assembly process is also affected by the hydrodynamic forces exerted on the colloid as they move along the surface in response to the external and inter-particle forces. For an isolated colloid, a hydrodynamic drag is exerted by the fluids bounding the interface as the colloid translates along the surface. Owing to the small size of the colloid and the strength of the inter-particle and external forces, the surface flows are typically inertialess (Stokes flow at low Reynolds number) and do not generate interfacial deformation (small capillary number). For isolated spherical particles, the drag coefficients for Stokes flows of particles moving over flat interfaces have been calculated and are a function of the immersion depth and the viscosity ratio of the bounding phases [58–64]. As colloids approach one another to a few particle radii, this hydrodynamic drag becomes strongly affected by the presence of the other particles. For the case of particle pairs, this *HI* contribution to the fluid forces exerted on the particle (again for inertialess flows and flat interfaces) can be divided into approaching/receding or in-tandem motions along the line of centers between the particles and oppositely directed (shear) or in-tandem motions along the axis perpendicular to the line of centers. These coefficients have also been calculated [65], and demonstrate that the *HI* associated with the approaching/receding or shearing relative motions correspond to large drag coefficients because of the viscous resistance of the fluid between the particles. As such, *HI* can be a strong modulator of the assembly process. As a model construct, colloidal monolayers are useful for studying particle motion in two dimensions under the action of external and internal forces. These systems are widely used to study various aspects concerning 2D particle dynamics such as self-organization under the influence of internal forces, ordered structure formation with short range and long range periodicity [66–70], crystal formation[71], phase changes [72] and two dimensional rheology [73,74,3,75,76]. The study of the effect of *HI* on the particle dynamics in this model system can lead to design rules of new particle-constructed materials that follow a bottom-up strategy [77,70] and take advantage of *HI*.

The incorporation of *HI* in the evolution of interfacial structure in 2D particle dynamics at a fluid interface has not been examined in detail. The aim of this study is to begin to understand this effect by undertaking Langevin dynamic simulations for the pairwise interaction of two particles straddling a flat fluid interface in Stokes flow, accounting for capillary attractive and charge repulsive inter-particle forces, Brownian motion and the hydrodynamic interaction. We briefly review the limited number of studies on the incorporation of *HI* in the 2D dynamics of particles at a fluid interface.

Examinations of the pairwise interaction of two spherical colloids attached to a planar fluid interface, and driven together by capillary attraction have been reported before [78,79,52,80,62]. These studies measured the center to center separation distance between a particle pair ($\ell(t)$) at various time steps. Additionally, the colloids are only allowed to translate along the interface since rotation is prohibited due to contact line pinning. The assumption stems from the fact that colloidal surfaces are rough at the micro scale and this causes contact line pinning on the particle surface as highlighted in numerous theoretical and experimental studies [81,63,82].

*HI* between colloids at fluid–fluid interfaces have been the focus of multiple previously published experimental and theoretical studies. Deghani and Barman et al. [83,84] imposed an external surface shear flow with a rotating Couette and reported theoretical results for a 2D Stokesian dynamics simulation of multiple colloids interacting at a fluid–fluid interface due to capillary attraction and electrostatic repulsion. This work was further developed in follow up studies that concentrated on self organization in the absence of an external flow field ([85–87]). The *HI* models included in these studies specifically incorporate the canonical modes 1 (Fig. 3 (Fig. 3(b)). Vidal et al. [88] numerically computed the Stokes drag on a planar array of colloids (density matched to the liquid to create a flat interface) on a gas/liquid interface ($d/a = 1$) by imposing a shear flow in a two dimensional channel, placing the particles at the center of the channel and then using the symmetry of the configuration to find the drag on the particles at the air/liquid interface as half the fully immersed drag. De and Huerre et al. [89,90] evaluated the nature of microstructure changes for particles attached to a planar or curved gas/liquid interface subject to periodic external actuation, and presented expressions for modified capillary attraction between these particles under normal rapid periodic forcing by considering inviscid flows.

Considerable research has been conducted on the 2D interfacial self-assembly of colloids driven by various types of forcing mech-anisms. Some studies have used Mesoscopic discrete element and Brownian dynamics simulations to set up a force balance between Stokes drag force on the particles, interparticle interactions and a stochastic Brownian force [91–95]. These studies utilize diffuse interface theory and factor in capillary attraction, electrostatic interaction forces, and frictional forces with a solid substrate (these studies focus on systems of particles embedded in a thin film of liquid placed on a solid surface) while modeling particle aggregation. However, these studies estimate hydrodynamic drag for colloidal systems by using the Stokes' drag formula for a single particle $6\pi\mu aU$, which does not include the influence of immersion depth into particle drag and HI stemming from particle pair interactions as highlighted in Figs. 3(a)–(d).

Using external magnetic fields to move particles for understanding interparticle interactions has become a popular strategy in recent times. These particles may be adsorbed at a gas/aqueous interface [57,96–98] or completely immersed in a liquid layer immediately above an inverted planar meniscus [99,100,68,69,101] or sandwiched between two parallel plates [102,103]. The particles interactions can easily be tuned as attractive or repulsive depending on direction of the applied magnetic field. Brownian dynamics simulations have been undertaken to estimate the diffusion of particles that are completely wetted and situated near to an inverted fluid interface and repel each other due to an applied normal magnetic field [99,100,69]. These studies apply pairwise HI expressions in an infinite medium by adopting a Rotne-Prager approach formulation [104], and present experimental data to prove that the diffusion coefficients of the colloids increases due to HI while interacting with each other via a repulsive $r^{-3}$ interparticle potential corresponding to the magnetic dipolar repulsion. Pesché et al.[105,106] have performed Stokesian Dynamics studies for colloidal monolayers of both charged and uncharged particles diffusing and interacting with each other with various interparticle potentials while the colloids remain trapped in a two dimensional plane situated exactly at the center of two parallel walls. They account for HI by using a combination of the drag coefficients for an infinite medium and the drag coefficient for the interaction of the particles with the wall. Although these studies concern particles immersed in a bulk fluid, they closely mimic the behavior for colloids at fluid–fluid interfaces when $d/a=1$ with the bounding media for the colloids being either a gas/liquid interface or a liquid/liquid interface with no viscosity contrast. Xie et al.[107,108] have used Lattice Boltzmann methods to model capillary and magnetic interactions of Janus particles on planar and drop interfaces under the influence of an externally applied magnetic field. Although the Lattice-Boltzmann method implicitly incorporates HI between the particles, explicit expressions that specifically address HI are not defined in such numerical approaches.

The purpose of this study is to provide a more exact treatment of HI on the pair-wise aggregation dynamics of Brownian colloids on a fluid interface and moving under the influence of capillary attraction, electrostatic repulsions and Brownian forces by undertaking Langevin (Brownian) dynamic simulations. We will consider two regimes for the pair interaction at the interface. In the first regime, the colloids are drawn together by a capillary attraction due to undulation of the contact line resulting from particles that are not smooth. Brownian forces, owing to their randomness, may or may not resist this attraction at an instance, and our objective will be to quantify the probability for aggregation and the mean time of aggregation through the Pe number, and the effect of the hydrodynamic interaction. In the second regime, the colloids are drawn together by an attraction (capillary attraction due contact line undulation) and resisted by a repulsive interaction (electrostatic repulsion due to charges on the colloid surface).

Electrostatic repulsion dominates at larger separations between the particles and capillary attraction dominates at smaller separations, creating a potential barrier to coalescence. The height of the barrier is a function of the scales of the electrostatic to the capillary attraction forces. The probability for aggregation is dependent on Brownian forces allowing the particles to scale this barrier to the smaller separations where capillary attraction can cause aggregation. Hence, Brownian forces play a more complicated role as they are necessary to overcome the barrier to aggregation, but at closer separations can cause particles not to aggregate under the dominance of the attractive force due to the random motion. For this regime we will again study how pair HI impact the probability for aggregation and the mean time for aggregation.

## 2. Problem Formulation

We consider the problem of two identical spherical colloids, radius $a$, straddling a planar fluid interface (Fig. 1) separating an upper nonpolar (e.g. air or oil) phase from a lower polar phase (e.g. an aqueous phase). We assume that the viscosity of the upper phase is negligible relative to the lower phase. At any instant in time, the particles are separated by a center-to-center distance $\ell$ and have velocities $\boldsymbol{U}_1$ and $\boldsymbol{U}_2$ with components along and perpendicular to the line of centers. We assume the fluid flow is inertialess (Stokes flow) and the interface is flat up to the surface of the floating particle (small capillary number). The particles are assumed to be neutrally wetting $\theta = 90^0$, and therefore the immersion depth $d/a$=1.

The particles are subject to the hydrodynamic drag forces, Brownian stochastic forces due to the thermal fluctuations of the solvent molecules surrounding the colloid ($\boldsymbol{F}_{Brownian}$), inter-particle capillary attractive forces ($\boldsymbol{F}_{Attraction}$, e.g. undulating contact line quadrapolar interactions as in Fig. 2)) and electrostatic repulsive forces ($\boldsymbol{F}_{Repulsion}$, e.g. the electrostatic repulsion between colloids with surface charge surrounded by a diffuse layers of counter ions of electrolyte in the lower polar phase as in Fig. 2 (b)). These forces are all column vectors with four rows; the first two rows are the forces on the first particle (denoted with vector $\boldsymbol{U}_1$) in the directions along and perpendicular to the line of centers, and the second two rows are the same quantities for the second particle. As we detail below, the inter-particle forces act along the line of centers of the particles, while the Brownian and hydrodynamic drag forces have components along and perpendicular to the line of centers. In this development, we assume that the particles do not rotate, principally because colloids are typically rough and chemically heterogeneous, and these effects cause pinning of the contact line (cf. the review [6]).

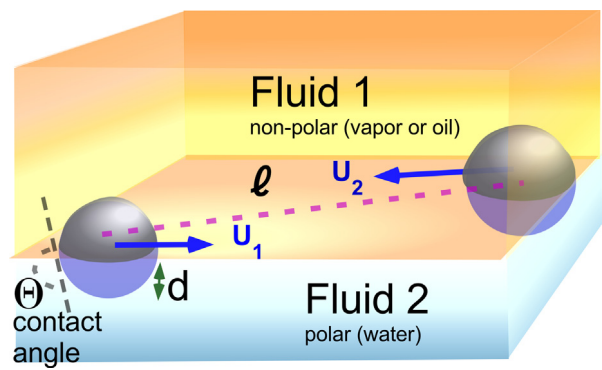

**Fig. 1.** A pictorial representation of a pair of particles of radius $a$ separated by a distance $\ell$ and moving with arbitrary velocities $U_1$ and $U_2$ along a flat interface between a upper hydrophobic phase (fluid 1 (air or oil)) and a lower hydrophilic liquid phase (fluid 2, e.g. water). The particles are neutrally wetting and the upper phase is much less viscous than the lower water phase.

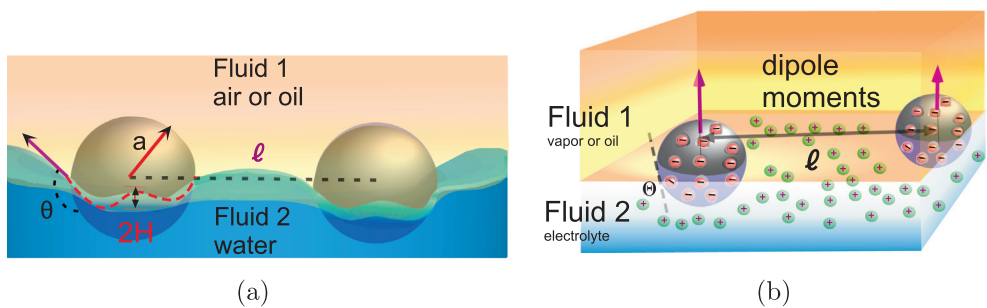

**Fig. 2.** (a) Capillary attraction of two particles of radius *a* at an interface between fluids 1 (air or oil) and 2(water) due to interfacial deformation contributed by an undulating contact line. (b) Electrostatic repulsion between charged particles at an air (or oil)/water interface.

To calculate particle trajectories along the interface which incorporate the effect of Brownian motion, hydrodynamic drag and inter-particle forces, we formulate an overdamped (vector) Langevin equation for the horizontal force balance on each of the particles:

$$\boldsymbol{F}_{Brownian} + \boldsymbol{F}_{Drag} + \boldsymbol{F}_{Attraction} + \boldsymbol{F}_{Repulsion} = 0 \quad (1)$$

where all forces lie on the plane of the interface. Expressions for the inter-particle capillary attractive forces, the electrostatic repulsive forces, the hydrodynamic drag forces and the explicit finite time difference form for the numerical integration of the Langevin equation are detailed in the following subsections.

*2.1. Contact Line Driven Capillary Attractive Forces*

For micron and sub-micron sized spherical colloids in which gravitational forces and gravity-driven capillary attractions are negligible (small Bond number), capillary attractions have been observed and attributed to surface topological heterogeneities (particle roughness) or chemical heterogeneities which pin the contact line in an undulating shape and form a rough meniscus around the colloid [40]. By assuming that the contact line undulations were small, and the deformation of the surrounding meniscus can be described by the linearized Young–Laplace equation, solutions for the capillary attraction were obtained by expanding the shape undulations and the accompanying deformation of the interface in a series expansion and computing the capillary force as a multipole sum. The energy of the quadrupolar interaction ($\Delta E$) is given by:

$$\Delta E = \frac{12\pi\gamma H^2 a^4 \sin^4\theta}{\ell^4} \quad (2)$$

where $\gamma$ is the interfacial tension between the two fluids, $\theta$ is the contact angle of the colloid at the fluid–fluid interface measured from the lower liquid phase and $H$ is the amplitude of the fluctuation of the contact line. The long range nature of the multipolar (quadrupolar) capillary attraction is evident from Eq. (2). For an inter-particle spacing of $\ell/a = 5$, the interaction energy ($\Delta E$) is comparable to the thermal energy ($kT$) of the particles for an undulation amplitude as low as $H \sim 0.5$ nm. For higher values of contact line undulation ($H \sim 100$ nm), the interaction energy greatly exceeds the thermal energy ($\Delta E \sim 10^4 kT$ for $\ell/a \sim 5$), which explains the high propensity of micron sized particles to cluster rapidly. For a pair of colloids interacting strictly under the influence of capillary quadrupoles, Eq. (2) can be differentiated to obtain the spacing dependent attractive force ($F_{quad}$) between the colloidal pair (magnitude given below).

$$F_{quad} = \frac{48\pi\gamma H^2 a^4 \sin^4\theta}{\ell^5} \quad (3)$$

The relative importance of the inter-particle capillary attractive interaction energy due to the contact line undulations relative to the thermal energy $kT$ can be scaled as $\frac{F_{quad}a}{kT} = \frac{48\pi\gamma H^2 a^5 \sin^4\theta}{kT\ell^5}$. The latter ratio defines a *Pe* number, which serves as a dimensionless quantity to understand the scaling of the deterministic forces to the diffusive forces. For a contact angle of 90°, a surface tension $\gamma$ of the interface characteristic of the air/water surface (72.6 mN/m), and a separation distance of $\ell/a = 5$ of the colloids from each other, altering the value of $H$ from a few nanometer to a sub micron scale modulates the value of *Pe* number from $0.01 - 100$ and can thereby result in almost five orders of magnitude increase in the quadrupolar interaction energy to the thermal energy.

*2.2. Electrostatic Repulsive Forces Due to Surface Charge*

When a charged colloid is placed on a fluid–fluid interface, the surface charges orient themselves with a different density in each of the adjoining fluid phases (see Fig. 2(b)). The density of the charges varies based on the dielectric constant of each phase. Such an asymmetric reorientation of charges at a fluid–fluid interface having a stratification with reference to the dielectric properties gives rise to a net dipole moment *p*. In case of a particle pair with both particles having an identical surface charge distribution, the dipoles on both particles can lead to a long range force of dipolar repulsion ($F_{repulsion}$) leading to a $\ell^4$ dependence which is given by

$$F_{repulsion} = \frac{3p^2}{2\epsilon_n\ell^4} \quad (4)$$

where $\epsilon_n$ is the dielectric constant of the non polar phase (air/oil). Since water is known to have a high dielectric constant ($\epsilon_{water} = 80$) as compared to standard non-polar oils ($\epsilon_n \sim 2 - 6$) or air, the surface charges present in the aqueous phase get screened by the counter ions thereby reducing the net dipole moment. However, the presence of residual surface charges on the non-polar side can still give rise to long range dipolar repulsion. Danov et al.[109] considered the case of a charged colloid, which is irreversibly adsorbed on a fluid–fluid interface and analytically obtained the electrical potential in the adjoining phases by solving a Laplace equation in both the phases. They assumed that the surface charge density on the non-polar side ($\sigma_n$) is known a priori and that the potential field is uniform in the aqueous phase because of its high dielectric constant. Additionally, they impose continuity of electric potential at the particle-nonpolar fluid interface and set the flux to be a function of the surface charge density. By applying a Mehler-Fock transform to solve the problem, they show that *p* (for particles in the colloidal size range and lower) can be represented as

$$p = 4\pi\sigma_n D\left(\frac{\epsilon_p}{\epsilon_n}, \theta\right) a^3 \sin^3\theta \tag{5}$$

where $\epsilon_p$ is the dielectric constant of the particle and $D$ is a dimensionless constant which is obtained by numerically solving a Fredholm integral equation of the second kind after performing a Mehler-Fock transform on the Laplacian of the potential field in the non-polar phase and it depends on the contact angle of the colloid and the ratio of the dielectric constants of the particle and the non polar phase. In order to speed up calculations, an interpolation curve is constructed using tabulated values of $D$ for various values of $\theta$ and ratios of $\epsilon_p/\epsilon_n$ and used in the code.

When the attractive forces generated due to capillary dipoles balance the dipolar repulsions due to surface charges, there is an equilibrium separation distance between particles ($\ell_{eq}$), where the net force acting on each particle is zero. In the absence of any particle motion, the forces between two particle can be given by:

$$F_{Attraction} + F_{Repulsion} = 0 \tag{6}$$

After substituting (4) and (3) into Eq. (6), we get

$$\frac{\ell_{eq}}{a} = \frac{2\epsilon_n \gamma H^2}{\pi \sigma_n^2 D^2 \sin^2\theta a^3} \tag{7}$$

The net potential energy of the particle pair $U = U_{Attraction} + U_{Repulsion}$ is given by the sum of the attractive and repulsive potentials which are given by:

$$U_{Attraction} = \frac{-12\pi\gamma H^2 a^4 \sin^4\theta}{\ell^4} \tag{8}$$

$$U_{Repulsion} = \frac{p^2}{2\epsilon_n \ell^3} \tag{9}$$

where $U_{Attraction}$ has a negative sign because of the attractive nature of the potential field. Using Eq. ()()()(7)–(9), it can be shown that the energy of the barrier $U_{bar}$ at the equilibrium distance $\ell_{eq}$ can be represented by:

$$\frac{U_{bar}}{kT} = \frac{\pi^5 \sigma_n^8 D^8 \sin^{12}\theta a^{12}}{4\epsilon_n^4 \gamma^3 H^6} \tag{10}$$

### 2.3. Hydrodynamic Forces

As noted earlier, the hydrodynamic motion of two particles moving along a planar fluid interface in the Stokes flow limit can be decomposed into four fundamental canonical motions[65] which is shown in detail in Fig. 3(a)–(d): 1) approaching each other with velocity $U$ along the line of center (Fig. 3(a)) 2) moving in tandem with velocity $U$ along the line of center (Fig. 3(b)) 3) moving in opposite direction with velocity of magnitude $U$ perpendicular to line of center (Fig. 3(c)) 4) moving in same direction with velocity of magnitude $U$ perpendicular to line of center(Fig. 3(d)). With each mode, there is a drag exerted on the particle in the flow direction which can be formulated in terms of nondimensional drag coefficients. In general, these drag coefficients are a function of the dimensionless immersion depth ($d/a$), dimensionless separation distance $\ell/a$ and the viscosity ratio of the fluids bounding the interface although here we restrict attention to an inviscid upper phase and neutral wetting ($d/a$=1). Thus for the modes along the line of centres, the dimensional drag on the particles along the center-to-center axis are given by $6\pi\mu a f_i(d/a, \ell/a) U$ (i = 1, 2) where $f_i(d/a, \ell/a)$ is the dimensionless drag coefficient (the drag is nondimensionalized by $6\pi\mu a U$). For the modes involving motion perpendicular to the line of centers, the dimensional drag is $6\pi\mu a f_i(d/a, \ell/a) U$ (i = 3, 4). Owing to the symmetry in the hydrodynamic motion for the neutrally wetting case, the surface drag is one-half the value in an infinite medium or $(1/2)\left[6\pi\mu U a \hat{f}_{i,\infty}(\ell/a)\right]$ where $\hat{f}_{i,\infty}(\ell/a)$ is the nondimensional drag coefficient in an infinite medium for motion $i$ and tends to one at infinite separation ($\lim_{\ell/a \to \infty} \hat{f}_{i,\infty}(\ell/a) = 1$) physically indicating no HI between particle pairs at large distances.

The solutions for the hydrodynamic drag coefficients in an infinite medium have been previously obtained analytically by using a bi-spherical coordinates formulation for two particles in a bulk fluid[52,110–112] and are listed below for reference.

$$\hat{f}_{1,\infty}\left(\frac{\ell}{a}\right) = \left\{1 + \frac{a}{2(\ell - 2a)}\right\}\left\{1 + 0.38 e^{-\{\ell n\frac{\ell}{a} + 0.6789\}^2/6.3}\right\} \tag{11}$$

$$\hat{f}_{2,\infty}\left(\frac{\ell}{a}\right) = \frac{4}{3}\sinh\alpha \times \\ \sum_{n=1}^{\infty}\left[\frac{n(n+1)}{(2n-1)(2n+3)} \times \left\{1 - \frac{4\sinh^2\left(n+\frac{1}{2}\right)\alpha - (2n+1)^2\sinh^2\alpha}{2\sinh(2n+1)\alpha + (2n+1)\sinh 2\alpha}\right\}\right] \tag{12}$$

$$\hat{f}_{3,\infty}\left(\frac{\ell}{a}\right) = 1 - \frac{3}{4}\left(\frac{a}{\ell}\right) + \frac{9}{16}\left(\frac{a}{\ell}\right)^2 \\ - \frac{59}{64}\left(\frac{a}{\ell}\right)^3 + \frac{465}{256}\left(\frac{a}{\ell}\right)^4 - \frac{15813}{7168}\left(\frac{a}{\ell}\right)^5 + \frac{2\left(\frac{a}{\ell}\right)^6}{1+\left(\frac{a}{\ell}\right)} \tag{13}$$

$$\hat{f}_{4,\infty}\left(\frac{\ell}{a}\right) = \frac{\sqrt{2}}{3}\sinh(\alpha)\sum_{n=0}^{\infty} D_n \tag{14}$$

Where $D_n$ are obtained as an exact solution of the problem in a bi-spherical coordinate system[112] and $\alpha$ is given by:

$$\alpha = \cosh^{-1}\left(\frac{\ell}{2a}\right) \tag{15}$$

The hydrodynamic drag force $F_{Drag}$ between a pair of colloidal particles straddling a fluid–fluid interface in this case can be given by

$$\mathbf{F}_{Drag} = \mathscr{R}\mathbf{U} \tag{16}$$

where $\mathbf{U}$ is a column vector of four components, consisting of the velocity along and perpendicular to the line of centers for particle "1", and particle "2" and $\mathscr{R}$ is the resistance tensor and given for neutral wetting particle pairs by the corresponding coefficients in an infinite medium:

$$\mathscr{R} = 6\pi\mu a \begin{pmatrix} \left(\frac{\hat{f}_{1,\infty}+\hat{f}_{2,\infty}}{4}\right) & \left(\frac{\hat{f}_{2,\infty}-\hat{f}_{1,\infty}}{4}\right) & 0 & 0 \\ \left(\frac{\hat{f}_{2,\infty}-\hat{f}_{1,\infty}}{4}\right) & \left(\frac{\hat{f}_{1,\infty}+\hat{f}_{2,\infty}}{4}\right) & 0 & 0 \\ 0 & 0 & \left(\frac{\hat{f}_{3,\infty}+\hat{f}_{4,\infty}}{4}\right) & \left(\frac{\hat{f}_{3,\infty}-\hat{f}_{4,\infty}}{4}\right) \\ 0 & 0 & \left(\frac{\hat{f}_{3,\infty}-\hat{f}_{4,\infty}}{4}\right) & \left(\frac{\hat{f}_{3,\infty}+\hat{f}_{4,\infty}}{4}\right) \end{pmatrix} \tag{17}$$

The detailed derivation of the resistance tensor $\mathscr{R}$ and definition of $\mathbf{U}$ and $F_{Drag}$ (the directional components of these variables are defined in a local coordinate system) can be found in A.

### 2.4. Brownian Forces and Langevin Equation

In order to compute the particle trajectories, Eq. (1) can be discretized to yield an evolution equation for the instantaneous position of each particle.

$$\Delta r = \mathscr{R}^{-1}(F_{Net})\Delta t + kT\nabla \cdot \mathscr{R}^{-1}\Delta t + r(\Delta t) \tag{18}$$

where $\Delta r$ is the instantaneous displacement vector of each particle, $\mathscr{R}^{-1}$ is the mobility tensor, $k$ is the Boltzmann constant, $T$ is the absolute temperature, $\mathbf{F}_{Net} = \mathbf{F}_{Attraction} + \mathbf{F}_{Repulsion}$ is the total deterministic force on the particle and $r(\Delta t)$ is a stochastic variable which governs the Brownian displacement of each particle. $\mathscr{R}^{-1}$ is a unique function of the center to center distance between the particles ($\ell$) and has to be computed at every time step ($\Delta t$). Since $r(\Delta t)$ is a stochastic variable, it satisfies

$$r(t) = \sqrt{2kT\Delta t}\sqrt{\mathscr{R}^{-1}} \cdot G(t) \qquad (19)$$

where $G(t)$ is a random normal variable with zero mean and unit variance.

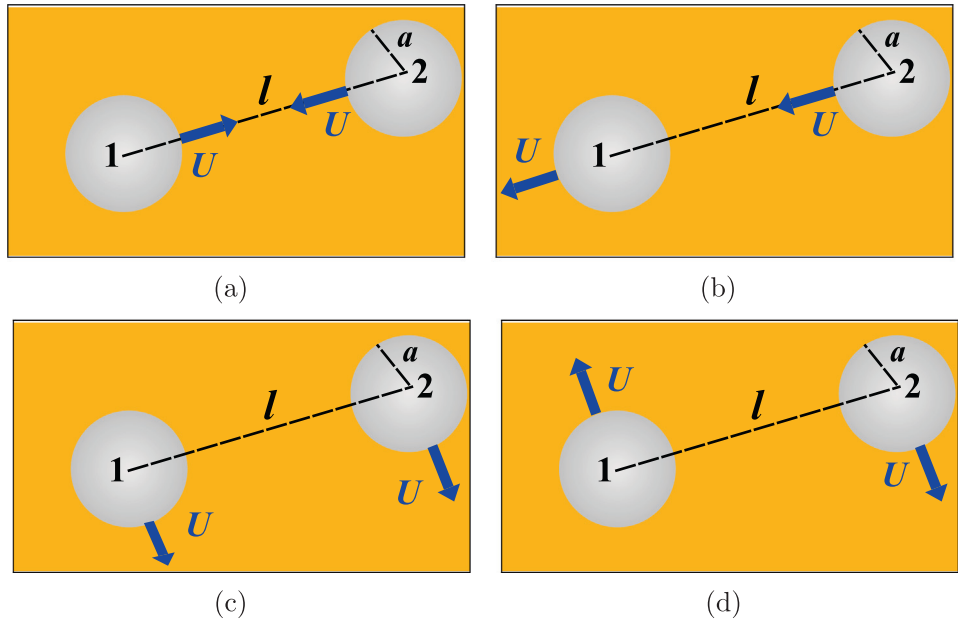

**Fig. 3.** A bird's eye view of two particles moving along a fluid–fluid interface with a velocity of magnitude $U$. The different modes of in plane hydrodynamic interactions are given by: a) Particle motion anti-parallel to each other and along their line of center ($f_1$) b) Particle motion parallel to each other and along their line of center ($f_2$) c) Particle motion parallel to each other and perpendicular to their line of center ($f_3$) d) Particle motion anti-parallel to each other and perpendicular to their line of center. ($f_4$).

$$\langle G(t) \rangle = 0 \qquad (20)$$

$$\langle G(t)G(t) \rangle = 1 \qquad (21)$$

## 3. Algorithm For Numerical Simulation Of Langevin Equation

In the absence of stochastic Brownian forces (only deterministic forces), $f_1$ and $f_2$ are dominant canonical motions because the relative motion of particles is only along their line of centres. However, the incorporation of thermal fluctuations adds a random component to particle motion in addition to the external force driven deterministic particle motion. Once the resistance tensor is calculated, the drift term ($kT\nabla \cdot \mathscr{R}^{-1}\Delta t$) in Eq. (18) can be estimated by using a mid-point algorithm[113]. In order to estimate the mean drift term, we assume that the initial positions on both the particles are given by $r_a$. We then consider Eq. (1) in the absence of any deterministic forces. In this case, the particle velocities ($u_{brow}$) are attributed to only the stochastic Brownian force ($F_{Brownian}$), which follows

$$\langle F_{Brownian}(t)F_{Brownian}(t) \rangle = \frac{2kT\mathscr{R}}{\Delta t} \qquad (22)$$

In the midpoint algorithm, the particles are displaced in a half time step ($\Delta t/2$) from a position $r_a$ to a position $r_b$ such that

$$r_b = r_a + u^a_{brow}\frac{\delta t}{2} \qquad (23)$$

All variables defined at position $r_a$ are denoted by subscript $a$ and all variables defined at position $r_b$ are denoted by subscript $b$ henceforth. Another principal assumption that the midpoint algorithm is that the stochastic forces on the particles at the initial and final position are the same ($F^a_{Brownian} = F^b_{Brownian}$). Eq. (1) can be written at the initial and final position as

$$\mathscr{R}^a u^a_{brow} = F^a_{brow} \qquad (24)$$

$$\mathscr{R}^b u^b_{brow} = F^b_{brow} \qquad (25)$$

Similarly, we can describe the derivative of the resistance tensor as

$$\nabla \cdot (\mathscr{R}^{-1}) = \frac{(\mathscr{R}^{-1})^b - (\mathscr{R}^{-1})^a}{r_b - r_a} \qquad (26)$$

It can now be easily shown that

$$kT\nabla \cdot (\mathscr{R}^{-1}) = \frac{(u^b_{brow} - u^a_{brow})\Delta t}{2} \qquad (27)$$

Calculation of the random displacement term ($r(\Delta t)$) requires computation $\sqrt{\mathscr{R}^{-1}}$, which is done by constructing a Cholesky decomposition of the resistance tensor.

Once the expressions for all the forces are well defined, the Langevin equation for each particle can be solved for different values of the Péclet number. For the case of purely attractive interactions, the relative importance of the deterministic force ($F_{Attraction}$) and the stochastic force ($F_{Brownian}$) can be captured by defining a *Pe* number (when the inter-particle separation $\ell$ equals 5a), which is given by

$$Pe = \left[\frac{F_{attraction}a}{kT}\right]_{\ell=5a} \qquad (28)$$

where $k$ is the Boltzmann constant and $T$ is the absolute temperature. Such an attractive force could be due to gravity induced capillary interactions for large particles or surface asperity induced capillary interactions for micron and sub-micron size particles. $\ell = 5a$ is a good starting point for studying particle aggregation dynamics since the capillary attractive force ($F_{Attraction}$) has a similar order of magnitude to the thermal Brownian force ($F_{Brownian}$) for values of $H$ as less as 0.5 nm.

It is obvious that there are two distinct regimes of motion to consider: 1) a high *Pe* number regime (where capillary attractive forces dictate the aggregation dynamics of the particle pairs) 2) a low Pe number regime (where Brownian diffusion dictates the aggregation dynamics of the particle pairs). Each regime has a different time scale depending on which force is dominant. In a high Pe number regime, the time scale is convective in nature (since it is dictated by capillary attractions) and defined as $\tilde{t} = 6\pi\mu a^2/(F_{attraction})_{\ell=5a}$. In a low Pe number regime, the time scale is diffusive in nature (since it is dictated by thermal fluctuations) and defined as $t^* = 6\pi\mu a^3/kT$. The typical time of simulation is chosen as ten times the appropriate (dominant) time scale for the simulation (for low Pe number, it is the diffusion time scale and for high Péclet number, it is the convective time scale). All simulations reported in this paper are seeded at an initial separation of five particle radii. To ensure accuracy, the normalized time step for each simulation ($\Delta\tilde{t}$ or $\Delta t^*$) is adjusted such that the instantaneous displacement of the particle in one time step does not exceed 5% of the particle radius ($\Delta r/a \leq 0.05$). In our case for all Pe numbers, total simulation time for any particle pair equals $5 \times 10^5 \times \Delta\tilde{t}$ or $5 \times 10^5 \times \Delta t^*$.

## 4. Results and Discussion

### 4.1. Interplay Of Pure Attraction And Thermal Interactions

Eq. (18) can be solved for the cases of different actuating forces depending on the size and the charge of the particles involved. Depending on the relative magnitude of the stochastic Brownian forces to the deterministic attractive or repulsive forces, aggregation of the particle pairs is not always guaranteed. Greater value of attractive forces improves the chances of particle aggregation and greater value of Brownian thermal force may decrease the chances of particle aggregation. In order to quantify the effect of Pe number on the probability of aggregation, further simulations were performed for multiple independent realizations (N being total number of realizations) at a fixed value of the Pe number. The particle pairs in each realization were initiated at the same starting point ($\ell/a = 5$) and the code was run for a reasonably long time (based on a convective or diffusive time scale as described earlier and as dictated by the Pe number). The total number of aggregated pairs ($N_{agg}$) at the end of simulation time were counted and the aggregation probability ($P_{aggregation}$) was computed as $P_{aggregation} = N_{agg}/N$. For the sake of statistical significance, N was chosen as 1000 for each Pe number. Simulations are conducted for sub-micron sized particles for $\theta = 90°$, interfacial tension $\gamma = 72 mN/m$, particle radius of 100 nm and contact line undulation ranging from $H = 0.2 - 2.8$ nm to change the Pe number.

Fig. 4(b) presents results for the probability of aggregation for particle pairs for $N = 1000$ with and without HI under the influence of capillary quadrupoles for various values of the Pe number (obtained by varying H). Regardless of the presence or absence of HI, $P_{aggregation}$ is qualitatively observed to reduce with a drop in the value of Pe. At higher values of the Pe number ($Pe > 1$), the red curve (HI present) and the blue curve (HI absent) both asymptote to a value of $P_{aggregation} = 1$. In other words, when deterministic attractive forces dominate significantly over the stochastic forces, particle aggregation is guaranteed regardless of the magnitude of HI thereby leading to a trivial result. When $Pe < 1$, the incorporation of HI changes $P_{aggregation}$ (by almost as much as 10% in certain instances depending upon the value of Pe) especially once the particles enter a Brownian force dominated regime. This highlights the growing importance of HI at micron and submicron length scales.

The effect of HI is observed more clearly upon a detailed inspection of the mean aggregation time ($t_{aggregation}$) and shape of the histogram for probability distribution function of the aggregation time (henceforth referred to as P.D.F) for each particle pair for $N = 1000$. The P.D.F value is obtained/plotted as follows: for a particular value of Pe, if x out of 1000 particle pairs aggregate, the aggregation time for each of the x particle pair is binned into 30 different equally spaced buckets of time. If y particle pairs out of the x particle pairs aggregate in a specific time bucket, the P.D.F value for that bucket is obtained using the simple relation $P.D.F = y/x$. $t_{aggregation}$ is calculated by obtaining the mean of the aggregation time recorded for all x pairs that aggregate (out of $N = 1000$) for a specific Pe number. For the case of the $Pe < 1$ (dominant Brownian motion), Fig. 4(a) presents the variation of the mean aggregation time ($t_{aggregation}$) with Pe number for particle pairs with and without HI. Regardless of HI, $t_{aggregation}$ always decreases with increasing Pe number and appears to converge for both cases at higher Pe. This is indicative that HI matter less when the deterministic component of motion increases. On the other hand, the high divergence of $t_{aggregation}$ between HI and no HI case at lower Pe number supports the hypothesis that greater HI between particles can substantially delay the onset of aggregation when driving forces for attraction are insignificant or comparable to stochastic forces. Inspection of the histograms of the P.D.F of $P_{aggregation}$ plotted against aggregation time with and without HI for the case of low Pe number (Fig. 5)) and high Pe number (Fig. 5(b)) confirms the importance of HI at lower Pe number. The P.D.F histogram for the high Pe number is right skewed with a long tail for the case of particles both with and without HI, which indicates that very few particles out of 1000 particle pairs considered in the simulation will not aggregate and validates that the simulation time scale choice is reasonable. The histogram for the low Pe number is right skewed with a long tail for the case of particles without HI (in blue color) and the same histogram appears more gaussian for particles with HI (in red color) thereby demonstrating the impact of including HI in predicting the aggregation behavior of particles as the length scale diminishes from sub-millimeter to nanometer dimensions. At values of low Pe number, the incorporation of HI further reduces the influence of attractive forces and that is the primary reason for the P.D.F vs aggregation time histograms to appear more gaussian in nature.

### 4.2. Effect Of Electrostatics On Particle Aggregation

When the particles deposited on a fluid interface are charged, an additional dipolar repulsive force ($F_{repulsion}$) exists between them. This force is long range in nature (see Eq. (4)) and decays at a rate lesser than capillary attractive forces dominant at these length scales($F_{repulsion} \sim \ell^{-4}$ and $F_{attraction} \sim \ell^{-5}$). As a result, repulsion dominates between the particles at large separations and attraction dominates between particles at smaller separations. There exists an equilibrium distance($\ell_{eq}$) between the particles when the attractive forces balance the repulsive forces leading to a net zero deterministic force. The exact value of $\ell_{eq}$ depends on the electrical and chemical properties of the particles along with their surface roughness (see Eq. (7)). The potential energy (U) of the particle pair is position dependent and depends on the sum of the attractive and repulsive potentials. By convention, we define the repulsive potential to be positive and the attractive potential to be negative. As a result, U is positive at large values of $\ell$ and negative at small values of $\ell$ and goes through a global maximum at $\ell = \ell_{eq}$. The maximum value of the potential ($U_{bar}$) is dictated by the electrical and chemical properties of the particles and is given by Eq. (10). All the simulations are initiated at an initial separation ($\ell_{start} = 5a$), which is situated in the repulsion dominated regime ($\ell_{start} > \ell_{eq}$). The potential at the starting position is given by $U_{start}$, which means that a particle pair has to overcome a potential barrier $\Delta U$ in order to aggregate (see Eq. (29)).

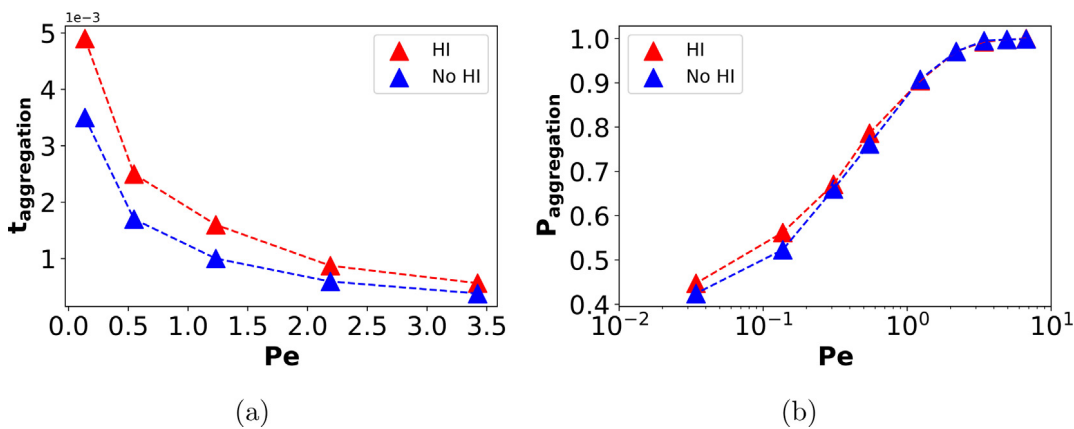

**Fig. 4.** Results for the case of uncharged particles a) Mean aggregation time in seconds with and without *HI* for different *Pe* numbers b) Aggregation probability for different *Pe* numbers.

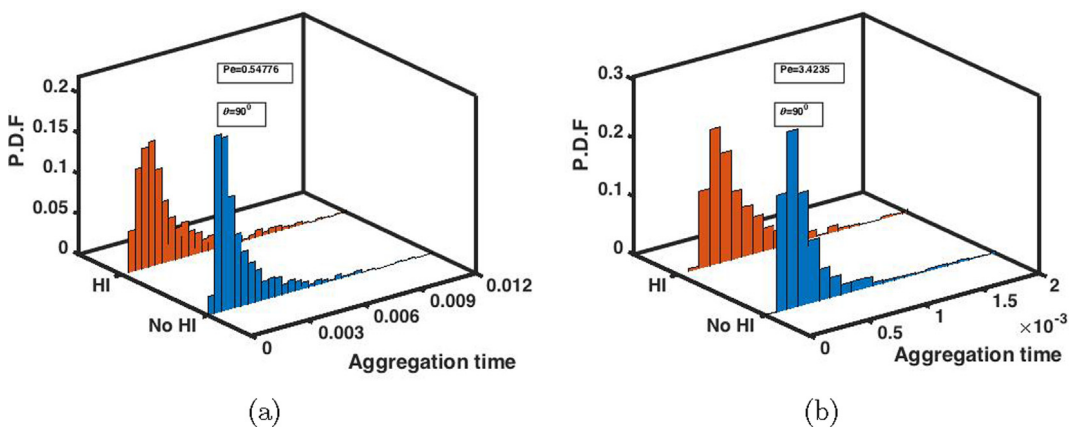

**Fig. 5.** Results for the case of uncharged particles a) 3D histogram comparing the aggregation time (P.D.F) in seconds for a particle pair for *HI* vs no *HI* when Brownian forces are dominant (*Pe* = 0.54776) b) 3D histogram comparing the aggregation time (P.D.F) in seconds for a particle pair for *HI* vs no *HI* when capillary attractive forces are dominant (*Pe* = 3.4235).

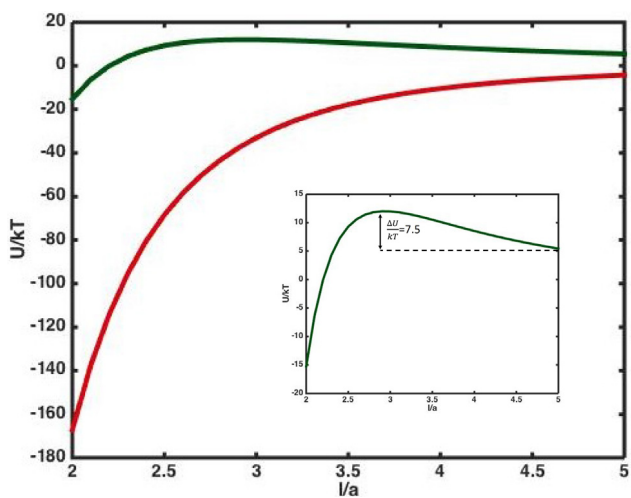

**Fig. 6.** Interaction potential between a pair of colloids as a function of the particle separation $\ell$. The red curve is the attractive potential for $\theta = 90°, a = 100$ nm and $H = 2$ nm. The green curve is the combined attractive and repulsive potential for $\theta = 90°, a = 100$ nm, $H = 2$ nm, $\epsilon_p/\epsilon_n = 4$ and $\Delta U/kT = 7.5$. The inset indicates the barrier $\Delta U$ encountered by the pair of particles in order to aggregate for the case of attractive and repulsive forces coexisting.

Fig. 6 provides an example of the interaction potential between particle pairs which attract each other due to capillary undulations and repel each other due to like electrostatic charges at the particle surface. In the absence of electrostatic repulsions (red curve), the interaction potential between the colloids decreases monotonically in magnitude with decreasing interparticle distance and eventually decays to zero at extremely large separations where interfacial deformation decays to zero. In the presence of both caplary attractive forces and electrostatic repulsive forces (green curve), the particle interaction potential has a negative value at small separation distance (the parameters chosen in this simulation correspond to a regime where attractive forces dominate over electrostatic repulsive forces at close proximity). As the distance between the particles increases, the influence of the attractive force decreases (due to a faster decay rate with $\ell$) and the influence of the repulsive force increases. This causes the magnitude of the interaction potential to reduce (note that the sign remains negative till attractive forces dominate) till we reach an equilibrium point where the capillary attractive and electrostatic repulsive forces equal each other. This corresponds to a local minimum in the absolute value of the interaction potential where $\ell$ equals $\ell_{eq}$. After the interparticle separation exceeds $\ell_{eq}$, electrostatic repulsions dominate and that reflects in the positive values of the interparticle potential.

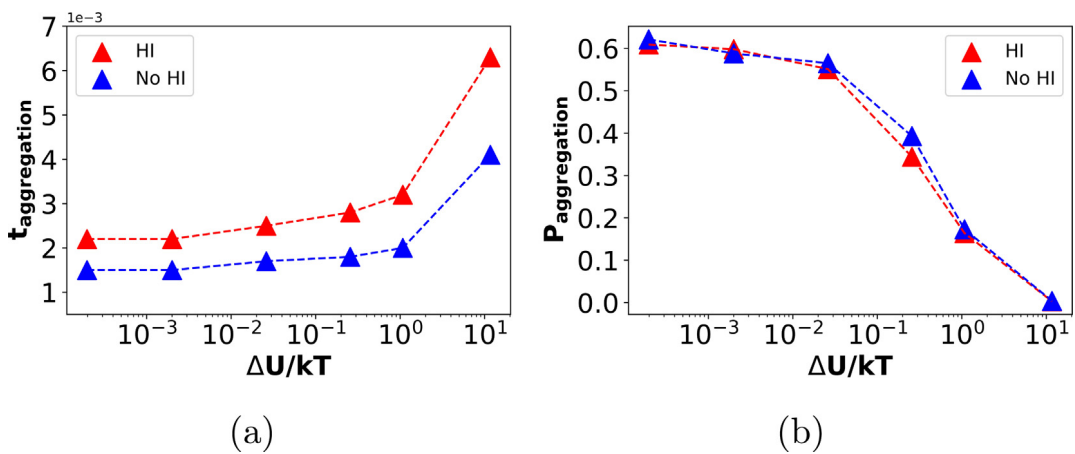

**Fig. 7.** Results for the case of charged particles a) Mean aggregation time in seconds with and without *HI* for different potentials b) Aggregation probability for different potentials (electrostatic repulsive forces).

Since both attractive and repulsive forces decay to zero at very large particle separations, the interaction potential decays to zero at very large separations.

$$\Delta U = U_{bar} - U_{start} \quad (29)$$

Simulations are conducted for micron and sub-micron sized particles for $\theta = 90°$, interfacial tension $\gamma = 72 mN/m$, a particle dielectric constant $\epsilon_p = 4$, a non-polar medium dielectric constant $\epsilon_n = 1$, surface charge densities ranging from $\sigma_n = 5.8 - 7.9 \times 10^{-5}$ C/m$^2$, particle radius of 100 nm and contact line undulation $H = 2$ nm. The surface charge density in the nonpolar phase $\sigma_n$ is varied from $\sigma_n = 5.8 - 7.9 \times 10^{-5}$ C/m$^2$ approximately based on previously reported experimental studies [47,114,115] and to change the intensity of the electrostatic repulsive force and thereby the height of the potential barrier ($\Delta U$). In order to focus attention on the interaction of particles under the influence of varying repulsive interactions, all simulations are conducted at a constant *Pe* number of 4.33 to ensure that the system has significant attractive forces as compared to the Brownian stochastic forces.

Fig. 7(b) presents results for the probability of aggregation for particle pairs for N = 1000 with and without *HI* for the same initial position (same *Pe* =4.33) and different inter-particle potentials which lead to different values of $\Delta U/kT$. When repulsive interactions are introduced, the absolute values of $P_{aggregation}$ are distinctly lower than those observed for purely attractive forces. A comparison of Fig. 4(b) and 7(b) shows that the value of $P_{aggregation}$ is greater than 0.9 for a *Pe* ~ 4.33 without electrostatic repulsion while the same value drops to almost 0.6 (a 30% decrease) with the introduction of weak electrostatic repulsions (values of $\Delta U/kT \sim 10^{-4}$). At extremely high values of $\Delta U/kT$, $P_{aggregation}$ asymptotically drops to zero for both *HI* present and *HI* absent cases thereby indicating that particles have no tendency to aggregate when repulsive forces are significantly greater than capillary attractive and Brownian forces. Some difference in $P_{aggregation}$ behavior is noted for charged particles in the presence/absence of *HI*. For uncharged particles, a comparison of $P_{aggregation}$ for any *Pe* number with *HI* present or absent indicates that $P_{aggregation}$ for the case of *HI* present either exceeds or equals $P_{aggregation}$ for the case of *HI* absent. For charged particles (as observed in Fig. 7(b)), $P_{aggregation}$ with *HI* present is slightly greater than $P_{aggregation}$ with *HI* absent in some cases and slightly lesser than $P_{aggregation}$ with *HI* absent in some cases. This difference in behavior is expected when repulsive forces are comparable to the attractive and Brownian forces in a system. When repulsive forces dominate, the natural tendency of the deterministic forces is to push the particles away from each other (as opposed to moving closer to each other when attractive forces dominate). Since the primary role of *HI* is to oppose any relative motion between the particles, *HI* now resist the motion of particles away from each other instead of towards each other.

The effect of *HI* is more pronounced upon inspection of the mean aggregation time and P.D.F of the particle pairs that manage to aggregate. Fig. 7(a) presents the variation of the mean aggrega-tion time for particle pairs with and without *HI* with $\Delta U/kT$ (note that a greater value of $\Delta U/kT$ corresponds to a higher value of repulsive force). It can be seen that $t_{aggregation}$ increases monotonically in general with increasing $\Delta U/kT$ whether *HI* are present or absent. For a fixed value of $\Delta U/kT$, $t_{aggregation}$ for the case of *HI* present exceeds the value of $t_{aggregation}$ for the case of *HI* absent (more by 60% in some cases). Histograms of the P.D.F. plotted against aggregation time with and without *HI* for the case of low repulsive forces (Figs. 8)) and high repulsive forces (Figs. 8(b)) confirms the importance of *HI*. The histograms for both the high repulsive force case and low repulsive force case appear slightly more Gaussian when *HI* are included. This is expected since *HI* dampen the contribution to particle displacement of deterministic forces by a factor of $\mathscr{R}^{-1}$ and of stochastic Brownian forces by a factor of $\mathscr{R}^{-0.5}$ as is evident from Eq. (18). With *HI* reducing the net displacement of the particles away from each other due to repulsive interactions, particles get more opportunity to drift near each other due to Stochastic Brownian motion which also increases the attractive force value slightly. The histograms for the *HI* absent case for both high repulsive and low repulsive force have relatively shorter tails as compared to the histograms for the *HI* present case. This is observed because *HI* increase the residence time of the particles in the vicinity of each other creating more chances for the particles to drift closer to each other via Brownian fluctuations (potentially into a zone where $F_{attraction}$ exceeds $F_{repulsion}$ and then allow attractive forces to dominate at closer range. The addition of *HI* is thereby critical in modeling particle behavior.

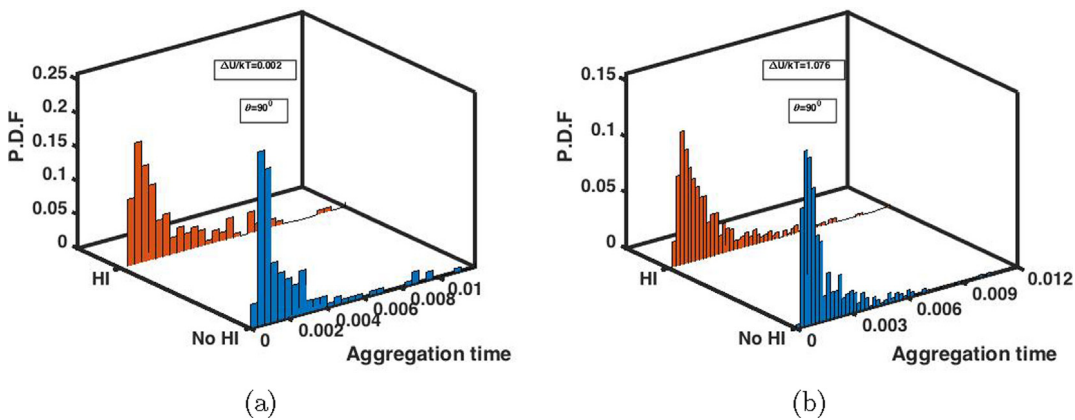

**Fig. 8.** Results for the case of charged particles a) 3D histogram (for a fixed capillary attractive force) comparing the P.D.F in seconds for a particle pair for *HI* vs no *HI* when Brownian forces are dominant over electrostatic repulsion ($\Delta U/kT = 0.002$) b) 3D histogram (for a fixed capillary attractive force) comparing the P.D.F in seconds for a particle pair for *HI* vs no *HI* when electrostatic repulsive forces are dominant over Brownian forces ($\Delta U/kT = 1.076$).

## 5. Conclusion

The clustering behavior of spherical particles at fluid–fluid interfaces (where the fluids adjoining the particles have a large viscosity contrast and the particles are equally wetted by the more viscous phase and less viscous phase) was studied by solving the Langevin equation for two particles. A model system comprising two identical particles trapped at a fluid–fluid interface is considered in the limit of low Reynolds Number and low Capillary number. The three phase contact line is assumed to be pinned on the surface of the particles based on experimental observations from previous studied[82].

Different inter-particle forces such as thermal fluctuations, capillary attractions (quadrupolar)[40] and electrostatic (dipolar) repulsion[109] were investigated in conjunction with defining a detailed mathematical framework that includes four different canonical modes of *HI*[65] stemming from particles being in the vicinity of each other. Explicit expressions for *HI*[52,110–112] that are derived by obtaining a bispherical solution of two particles interacting in bulk are used to compute the drag force while constructing the Langevin Dynamics equation. For the case of particles with contact angle $\theta = 90^0$, symmetry allows these explicit expressions for *HI* to be used directly without any adjustments or approximations. While the influence of *HI* on the aggregation of particle pairs adsorbed at fluid–fluid interfaces has been studied for sub-millimeter and micron sized particles before [78,79,52,80,62,82,85–87], these studies model the hydrodynamic drag between the particles either as Stokes drag for a single particle in bulk ($6\pi\mu a U$) or as Stokes drag for a particle pair moving towards each other on the interface along their line of center (only one canonical mode which leads to a drag force of $6\pi\mu a f_1(d/a, \ell/a)U$). We seek to address this gap by incorporating a total of four canonical modes of motion which account for drag forces generated from all four possible directions of relative motion between particle pairs at the fluid/fluid interface.

For the case of purely attractive interactions between particles, it was observed that the probability of forming aggregates scales with the magnitude of the actuating forces relative to the thermal interactions. When pure attraction exists between the particles (for the case of uncharged particles), the probability of aggregation of the particles decreases when *HI* are included for all values of the *Pe* number except for $\mathcal{O}(100)$ and above values. Incorporation of *HI* has a significant influence on mean aggregation time especially at low *Pe* numbers. While mean aggregation times are higher with *HI* particularly for low *Pe* numbers, the difference between the mean aggregation times gradually diminishes upon increasing *Pe* number and becomes nearly identical at high *Pe* number.

When charges exist on the particle surface, there is a unique value of the inter-particle distance ($\ell_{eq}$) where the net deterministic force on the particles is zero ($F_{Attraction} + F_{Repulsion} = 0$). The inter-particle potential at this point corresponds to a peak ($U_{bar}$) value, which plays an important role in determining aggregation probability. At distances greater than $\ell_{eq}$, the repulsive forces between the particles dominate and for distances smaller than $\ell_{eq}$, the attractive forces between the particles dominate. For same *Pe* number (same attractive force) and higher values of $\Delta U/kT$ (which signifies that the location of $U_{bar}$ is further distance wise from the starting point of the simulation), the probability of aggregation decreases because of the inability of the thermal interactions to move the particles across the energy barrier (located at $\ell = \ell_{eq}$) into the attractive zone in general. While presence of *HI* consistently reduces $P_{aggregation}$ for all values of *Pe* number for uncharged particles, the *HI* between the particles can marginally increase the probability of the aggregation for charged particles for certain values of $\Delta U/kT$. This anomalous behavior stems from the fact that *HI* fundamentally slow down any relative motion of particles and while relative motion of the particles is towards each other in an attraction dominated regime, the relative motion of the particles is away from each other in a repulsion dominated regime.

The theoretical concepts defined in our Langevin dynamics simulations and our results which demonstrate the correlation of *HI* to the probability of aggregate formation at fluid–fluid interfaces have broad interest spanning diverse fields such as pollination [11], bacteria motility modeling [12,116], process improvement in steel manufacturing[117], novel materials development through pickering emulsion stabilization[118] and determing 2-D self-assembly mechanisms that govern long range pattern formation required in semiconductor and environmental applications[119]. Comprehensive details on applications where understanding of *HI* is critical in improving our understanding of self-assembly processes can be found elsewhere[119,120]. This model and the results from these simulations can serve as a basis to provide a more quantitative interpretation of experimental data published in work focusing on cluster formation of micro-particles at oil–water interfaces[121] and measurement of attractive[122] and repulsive particle interactions[123] on an oil/water and air/water interface. Pérez-Juárez et al.[121], in their recently reported work, deposit monolayers of silica microspheres surface functionalized with methyl groups on a decane/water interface where decane and water are chosen because they have equal viscosity. Although contact angle is not reported, it is reasonable to expect the contact angle for these particles on such a decane/water interface to be around $90^0$ or greater. Using these assumptions, the diffusion coefficient for an isolated particle in their system as calculated by the Stokes–Einstein equation $kT/6\pi\mu a$ is approximately $0.152 \mu m^2/s$, while the diffusion coefficient measured in their experiments at low packing fraction (as seen in Fig. 4 of their paper) is approximately $0.13 \mu m^2/s$. This 17% difference between experimental data and theoretical data can be reduced if a modified theoretical diffusion coefficient is estimated by using the four canonical modes of *HI* described in our work. Aveyard et al.[124], in their experimental work, use laser tweezers as a tool to measure repulsion forces between charged polystyrene particles with diameter $2.7 \mu m$ at an oil–water interface where the oil phase is a mixture of decane and undecane blended to match the viscosity of water. The estimated contact angle of particles at interface is $70^0$. The authors use an optical tweezers setup to capture two particles straddling an interface in an optical trap by applying a known laser power $P$. They use a horizontal force balance of optical force (which is a function of $P$) and repulsive force to measure the repulsive force. In order to obtain a relationship between $P$ and laser force, they perform an independent calibration experiment where they place two particle at a center to center separation of $19 \mu m$ ($\ell/a \sim 14$) and move the stage perpendicular to the line of center of particle motion. This perpendicular motion of the stage generates a viscous resistive force on each particle and the authors record the velocity of the stage at which the particles can escape the optical trap and equate the force generated by the known laser power $P$ with Stokes drag $6\pi\mu aU$. The authors specify in their paper that actual drag expected in their experimental setup would be larger but still observe that using the Stokes drag expression as an approximation yields a measurement error of less than 10%. Our theoretical model provides a better insight into why this experimental result of the authors is accurate despite the approximations made. For the case of a single particle with a contact angle of $70^0$ straddling an oil–water interface, Pozrikidis et al.[125] have previously used Boundary integral methods to calculate that the drag force due to the presence of an interface to be around $\sim 1.1 \times 6\pi\mu aU$. Our model suggests that the drag force on two hydrodynamically interacting particles on such an interface needs to be $6\pi\mu af_3(d/a, \ell/a)U \sim 6\pi\mu a \times 1.1 \times f_{3,\infty}(\ell/a)U$. At a value of $\ell/a \sim 14, f_{3,\infty}(\ell/a) \sim 0.949$, which leads to $6\pi\mu af_3(d/a, \ell/a)U \sim 6\pi\mu a \times 1.1 \times 0.949 \times U \sim 0.95 \times 6\pi\mu aU$, which is less than 10% deviation as mentioned by the authors.

Future work for enhancing this model presents the following possibilities: 1) Particle aggregation for contact angles apart from $90^0$ can be explored further and quantified similarly by obtaining aggregation probability and mean aggregation time as a function of $Pe$ number for purely attractive particles and as a function of $\Delta U/kT$ for particles with attractive and repulsive forces. From a physical sense, two distinct regimes are expected to emerge for hydrophilic and hydrophobic particles. This difference is because the drag coefficients $f_1(d/a, \ell/a)$ and $f_3(d/a, \ell/a)$ increase significantly at close separation for hydrophilic particles and do not increase significantly for hydrophobic particles as shown through finite element simulations performed by Das et al.[65]. 2) Estimating the influence of a deformed interface (either due to gravity, contact line undulations or electrodipping force) on the motion of a single particle has become the recent focus of numerous recent studies. Dorr et al.[81] have used matched asymptotic expansions to calculate the flow field for a single translating sphere on a fluid–fluid interface with a huge viscosity contrast for $\mathcal{O}(0)$ Capillary number and observed that the flow velocity field generates a capillary force between two such spheres with an azimuthal angle dependence and an interparticle distance dependence that scales similar to capillary attractions between particles having an undulated contact line. Loudet et al.[126] performed finite element simulations for a 2-D cylindrical particle deposited on a fluid–fluid interface and exposed to a uniform velocity flow field across the interface for various contact angles, viscosity ratios and solid particle densities. Their calculations revealed a complex interplay between drag coefficients, viscosity ratio, contact angle and particle densities with anomalous drag coefficients obtained for the cases of finite viscosity ratios and particles that are not neutrally buoyant. Ben'MBarek et al.[127] performed free energy minimization calculations for polystyrene spheres of various sizes on a clean air–water interface to obtain drag coefficients as a function of par-
ticle deformation and showed that a curved meniscus can contribute to a nearly 42% increase in drag for a single particle at a fluid–fluid interface. All the currently published work on *HI* between particle pairs at a fluid interface assumes the existence of a flat interface around the particles since it is not possible to find a closed form solution if the fluid–fluid interface is allowed to deform in the presence of the particle. There is potential to use Boundary integral, finite element, Lattice-Boltzmann, molecular dynamics or any other appropriate numerical techniques to compute hydrodynamic drag experienced by a pair of spheres straddling an interface when both spheres generate a finite deformation of the interface.

**Declaration of Competing Interest**

The authors declare that they have no known competing financial interests or personal relationships that could have appeared to influence the work reported in this paper.

**Appendix A. Derivation of resistance tensor**

In order to obtain the generic resistance tensor for a pair of particles, we begin with the situation depicted by Fig. 9(a). The x axis is assumed to be the line joining the centers of each particle and the y axis is the positive normal to the x axis (see Fig. 9(a)). Let the x component of the velocity of particle 1 be $U_{1x}$ and the x component of velocity of particle 2 be $U_{2x}$. Similarly, the velocities of particle 1 and 2 along the y direction can be given as $U_{1y}$ and $U_{2y}$ respectively. Since drag coefficients are analytically known for well defined canonical modes with particles having equal velocity magnitude (Figs. 9(d)–(g)), the velocities of each particle ($U_{1x}, U_{2x}, U_{1y}$ and $U_{2y}$) have to be further resolved into four components (one component for every mode). For the sake of the derivation, we consider a snapshot of time where the configuration of the particles at the interface is fixed. For all the four modes under consideration, the particles are assumed to have a velocity of equal magnitude ($U_a, U_b, U_c$ and $U_d$ in Figs. 9(d)–(g)). On resolving the velocities along the x axis, we get

$$U_b = \frac{U_{x1} + U_{x2}}{2} \quad (A.1)$$

$$U_a = \frac{U_{x2} - U_{x1}}{2} \quad (A.2)$$

Similarly, resolution of the velocities along the y axis yields

$$U_c = \frac{U_{y1} + U_{y2}}{2} \quad (A.3)$$

$$U_d = \frac{U_{y2} - U_{y1}}{2} \quad (A.4)$$

In order to compute the drag forces associated with each mode of relative motion, it is necessary to define a drag coefficient separately for each case. The drag coefficients are a unique function of the configuration ($\ell/a$) of the particles and are previously defined in Eqs. (11)–(14). Note that Eqs. (11)–(14) are defined for particle pairs in a bulk fluid. For the specific case of a particle pair of contact angle $\theta = 90^0$ on a fluid/fluid interface with a high viscosity contrast (one fluid with high viscosity and the other fluid with vanishingly low viscosity), each drag coefficient $f_1, f_2, f_3$ and $f_4$ is required to be divided by two since only half the particle that is wetted by the more viscous phase experiences hydrodynamic drag and hydrodynamic drag force in the fluid phase with vanishing viscosity is negligible.

Once the drag coefficients and the corresponding velocities for each mode of relative motion are defined, the hydrodynamic drag on particle 1 in the x direction ($F_1^x$) and the y direction ($F_1^y$) and the hydrodynamic drag on particle 2 in the x direction ($F_2^x$) and the y direction ($F_2^y$) are given by

$$F_1^x = -6\pi\mu a\left(\frac{f_{1,\infty}}{2}\right)U_a + 6\pi\mu a\left(\frac{f_{2,\infty}}{2}\right)U_b$$
$$= 6\pi\mu a U_{1x}\left(\frac{f_{1,\infty}+f_{2,\infty}}{4}\right) + 6\pi\mu a U_{2x}\left(\frac{f_{2,\infty}-f_{1,\infty}}{4}\right) \quad \text{(A.5)}$$

$$F_2^x = 6\pi\mu a\left(\frac{f_{1,\infty}}{2}\right)U_a + 6\pi\mu a\left(\frac{f_{2,\infty}}{2}\right)U_b$$
$$= 6\pi\mu a U_{1x}\left(\frac{f_{2,\infty}-f_{1,\infty}}{4}\right) + 6\pi\mu a U_{2x}\left(\frac{f_{1,\infty}+f_{2,\infty}}{4}\right) \quad \text{(A.6)}$$

$$F_1^y = 6\pi\mu a\left(\frac{f_{3,\infty}}{2}\right)U_c + 6\pi\mu a\left(\frac{f_{4,\infty}}{2}\right)U_d$$
$$= 6\pi\mu a U_{1y}\left(\frac{f_{3,\infty}+f_{4,\infty}}{4}\right) + 6\pi\mu a U_{2y}\left(\frac{f_{3,\infty}-f_{4,\infty}}{4}\right) \quad \text{(A.7)}$$

$$F_2^y = 6\pi\mu a\left(\frac{f_{3,\infty}}{2}\right)U_c - 6\pi\mu a\left(\frac{f_{4,\infty}}{2}\right)U_d$$
$$= 6\pi\mu a U_{1y}\left(\frac{f_{3,\infty}-f_{4,\infty}}{4}\right) + 6\pi\mu a U_{2y}\left(\frac{f_{3,\infty}+f_{4,\infty}}{4}\right) \quad \text{(A.8)}$$

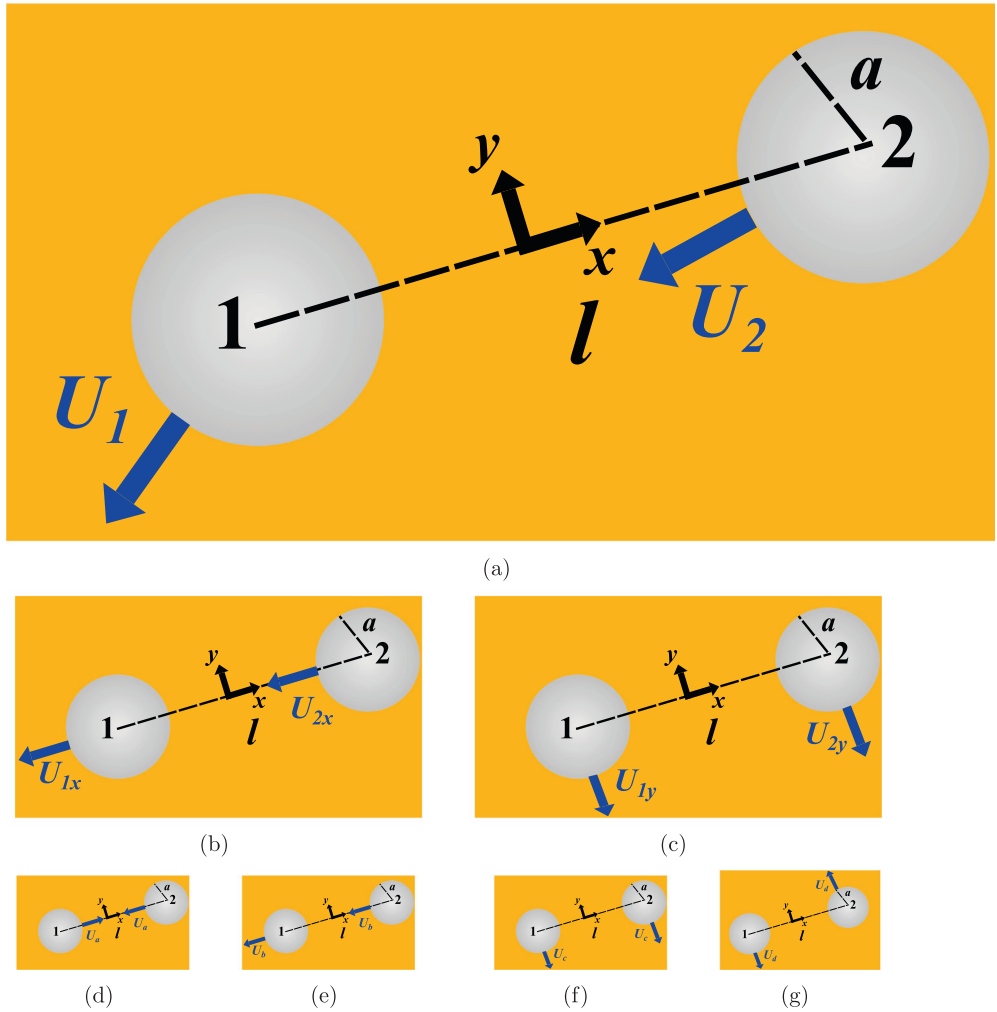

(a)

(b)     (c)

(d)     (e)     (f)     (g)

**Fig. 9.** Resolution of the main problem into relative motion of particles along (x axis) and perpendicular (y axis) to their line of center. The velocities are further resolved into the four fundamental modes of motion for which drag coefficients are well known. a) Top view of two particles moving with unknown velocities $U_1$ and $U_2$ in the interfacial plane b) Components of particle motion ($U_{1x}$ and $U_{2x}$) along the x (capillary) axis c) Components of particle motion ($U_{1y}$ and $U_{2y}$) along the y (normal) axis d) Aparallel relative motion of particles (relative to line of center) e) Parallel relative motion of particles (relative to line of center) f) Perpendicular relative motion of particles (relative to line of center) g) Aperpendicular relative motion of particles (relative to line of center).

Eqs. (A.5)–(A.8) can be represented as

$$\begin{pmatrix} F_1^x \\ F_2^x \\ F_1^y \\ F_2^y \end{pmatrix} = \mathscr{R} \begin{pmatrix} U_{1x} \\ U_{2x} \\ U_{1y} \\ U_{2y} \end{pmatrix} \quad \text{(A.9)}$$

where $\mathscr{R}$ is the resistance tensor compiled by collecting the coefficients of $U_{1x}, U_{2x}, U_{1y}$ and $U_{2y}$ and given by

$$\mathscr{R} = 6\pi\mu a \begin{pmatrix} \left(\frac{\hat{f}_{1,\infty}+\hat{f}_{2,\infty}}{4}\right) & \left(\frac{\hat{f}_{2,\infty}-\hat{f}_{1,\infty}}{4}\right) & 0 & 0 \\ \left(\frac{\hat{f}_{2,\infty}-\hat{f}_{1,\infty}}{4}\right) & \left(\frac{\hat{f}_{1,\infty}+\hat{f}_{2,\infty}}{4}\right) & 0 & 0 \\ 0 & 0 & \left(\frac{\hat{f}_{3,\infty}+\hat{f}_{4,\infty}}{4}\right) & \left(\frac{\hat{f}_{3,\infty}-\hat{f}_{4,\infty}}{4}\right) \\ 0 & 0 & \left(\frac{\hat{f}_{3,\infty}-\hat{f}_{4,\infty}}{4}\right) & \left(\frac{\hat{f}_{3,\infty}+\hat{f}_{4,\infty}}{4}\right) \end{pmatrix}$$

(A.10)

## References

[1] B. Binks, T. Horozov, Colloid paticles at liquid interfaces, in: Colloidal Particles at Liquid Interfaces, Cambridge University Press, 2006.
[2] V. Garbin, J. Crocker, K. Stebe, Nanoparticles at fluid interfaces: Exploiting capping ligands to control adsorption, stability and dynamics, J. Colloid Interface Sci. 387 (2012) 1–11.
[3] O.S. Deshmukh, D. van den Ende, M.C. Stuart, F. Mugele, M.H. Duits, Hard and soft colloids at fluid interfaces: Adsorption, interactions, assembly & rheology, Adv. Colloid Interface Sci. 222 (2015) 215–227.
[4] A. Maestro, E. Santini, E. Guzmán, Physico-chemical foundations of particle-laden fluid interfaces, Eur. Phys. J. E 41 (8) (2018) 97.
[5] E. Guzmán, F. Martínez-Pedrero, C. Calero, A. Maestro, F. Ortega, R.G. Rubio, A broad perspective to particle-laden fluid interfaces systems: From chemically homogeneous particles to active colloids, Adv. Colloid Interface Sci. (2022) 102620.
[6] C. Maldarelli, N.T. Donovan, S.C. Ganesh, S. Das, J. Koplik, Continuum and molecular dynamics studies of the hydrodynamics of colloids straddling a fluid interface, Annu. Rev. Fluid Mech. 54 (2022) 495–523.
[7] V. Garbin, J.C. Crocker, K.J. Stebe, Forced desorption of nanoparticles from an oil–water interface, Langmuir 28 (3) (2012) 1663–1667.
[8] S. Razavi, K.D. Cao, B. Lin, K.Y.C. Lee, R.S. Tu, I. Kretzschmar, Collapse of particle-laden interfaces under compression: buckling vs particle expulsion, Langmuir 31 (28) (2015) 7764–7775.
[9] V. Poulichet, V. Garbin, Ultrafast desorption of colloidal particles from fluid interfaces, Proc. Nat. Acad. Sci. 112 (19) (2015) 5932–5937.
[10] S.E. Anachkov, I. Lesov, M. Zanini, P.A. Kralchevsky, N.D. Denkov, L. Isa, Particle detachment from fluid interfaces: theory vs. experiments, Soft Matter 12 (36) (2016) 7632–7643.
[11] P.A. Cox, R.B. Knox, Two-dimensional pollination in hydrophilous plants: convergent evolution in the genera halodule (cymodoceaceae), halophila (hydrocharitaceae), ruppia (ruppiaceae), and lepilaena (zannichelliaceae), American journal of botany (1989) 164–175.
[12] S. Gart, D. Vella, S. Jung, The collective motion of nematodes in a thin liquid layer, Soft Matter 7 (6) (2011) 2444–2448.
[13] O. Velev, K. Furusawa, K. Nagayama, Assembly of latex particles by using emulsion droplets as templates. 1. microstructured hollow spheres, Langmuir 12 (10) (1996) 2374–2384.
[14] T.S. Horozov, B.P. Binks, Particle-stabilized emulsions: A bilayer or a bridging monolayer?, Angew Chem. 118 (5) (2006) 787–790.
[15] N. Bowden, A. Terfort, J. Carbeck, G.M. Whitesides, Self-assembly of mesoscale objects into ordered two-dimensional arrays, Science 276 (5310) (1997) 233–235.
[16] F. Burmeister, C. Schafle, T. Matthes, M. Bahmisch, J. Boneberg, P. Leiderer, Colloid monolayers as versatile lithographic masks, Langmuir 13 (11) (1997) 2983–2987.
[17] B. Prevo, D. Kuncicky, O. Velev, Engineered deposition of coatings from nano- and micro-particles: a brief review of convective assembly at high volume fraction, Colloids and Surfaces A -Physicochemical and Engineering Aspects 311 (2007) 2–10.
[18] A. Law, D. Buzza, T. Horozov, Two dimensional colloidal alloys, Phys. Rev. Lett. 106 (2011) 128302.
[19] S. Tarimala, L.L. Dai, Structure of microparticles in solid-stabilized emulsions, Langmuir 20 (9) (2004) 3492–3494.
[20] H. Ma, L.L. Dai, Structure of multi-component colloidal lattices at oil- water interfaces, Langmuir 25 (19) (2009) 11210–11215.
[21] R. Aveyard, J.H. Clint, D. Nees, N. Quirke, Structure and collapse of particle monolayers under lateral pressure at the octane/aqueous surfactant solution interface, Langmuir 16 (23) (2000) 8820–8828.
[22] R. Aveyard, J.H. Clint, D. Nees, V. Paunov, Compression and structure of monolayers of charged latex particles at air/water and octane/water interfaces, Langmuir 16 (2000) 1969–1979.
[23] T.S. Horozov, R. Aveyard, J.H. Clint, B.P. Binks, Order- disorder transition in monolayers of modified monodisperse silica particles at the octane- water interface, Langmuir 19 (7) (2003) 2822–2829.
[24] T. Horozov, R. Aveyard, B.P. Binks, J.H. Clint, Structure and stability of silica particle monolayers at horizontal and vertical octane-water interfaces, Langmuir 21 (2005) 7405–7412.
[25] S. Reynaert, P. Moldenaers, J. Vermant, Control over colloidal aggregation in monolayers of latex particles at the oil-water interface, Langmuir 22 (2006) 4936–4945.
[26] A.D. Law, M. Auriol, D. Smith, T.S. Horozov, D.M.A. Buzza, Self-assembly of two-dimensional colloidal clusters by tuning the hydrophobicity, composition, and packing geometry, Physical review letters 110 (13) (2013) 138301.
[27] L.L. Dai, R. Sharma, C.-Y. Wu, Self-assembled structure of nanoparticles at a liquid- liquid interface, Langmuir 21 (7) (2005) 2641–2643.
[28] S. Tarimala, C.-Y. Wu, L.L. Dai, Dynamics and collapse of two-dimensional colloidal lattices, Langmuir 22 (18) (2006) 7458–7461.
[29] C.-Y. Wu, S. Tarimala, L.L. Dai, Dynamics of charged microparticles at oil-water interfaces, Langmuir 22 (5) (2006) 2112–2116.
[30] F. Bresme, M. Oettel, Nanoparticles at fluid interfaces, J. Phys.: Condens. Matter 19 (2007) 413101.
[31] M. Nicolson, The interaction between floating particles, in: Mathematical Proceedings of the Cambridge Philosophical Society, Vol. 45, Cambridge University Press, 1949, pp. 288–295.
[32] D. Chan, J. Henry Jr, L. White, The interaction of colloidal particles collected at fluid interfaces, J. Colloid Interface Sci. 79 (2) (1981) 410–418.
[33] A. Wurger, Capillary attraction of charged particles at a curved interface, Europhys. Lett. 75 (2006) 978–984.
[34] C. Blanc, D. Fedorenko, M. Gross, M. In, M. Abkarian, M.A. Gharbi, J.-B. Fournier, P. Galatola, M. Nobili, Capillary force on a micrometric sphere trapped at a fluid interface exhibiting arbitrary curvature gradients, Physical review letters 111 (5) (2013) 058302.
[35] P. Galatola, J. Fournier, Capillary force acting on a colloidal particle floating on a deformed interface, Soft Matter 10 (2014) 2197–2212.
[36] P.A. Kralchevsky, V.N. Paunov, I.B. Ivanov, K. Nagayama, Capillary meniscus interaction between colloidal particles attached to a liquid-fluid interface, J. Colloid Interface Sci. 151 (1) (1992) 79–94.
[37] P.A. Kralchevsky, V.N. Paunov, N.D. Denkov, I.B. Ivanov, K. Nagayama, Energetical and force approaches to the capillary interactions between particles attached to a liquid-fluid interface, J. Colloid Interface Sci. 155 (2) (1993) 420–437.
[38] P.A. Kralchevsky, K. Nagayama, Capillary forces between colloidal particles, Langmuir 10 (1) (1994) 23–36.
[39] P.A. Kralchevsky, K. Nagayama, Capillary interactions between particles bound to interfaces, liquid films and biomembranes, Adv. Colloid Interface Sci. 85 (2000) 145–192.
[40] D. Stamou, C. Duschl, D. Johannsmann, Long-range attraction between colloidal spheres at the air-water interface: The consequence of an irregular meniscus, Phys. Rev. E 62 (4) (2000) 5263.
[41] K.D. Danov, P.A. Kralchevsky, Capillary forces between particles at a liquid interface: General theoretical approach and interactions between capillary multipoles, Adv. Colloid Interface Sci. 154 (1–2) (2010) 91–103.
[42] J. Lucassen, Capillary forces between solid particles in fluid interfaces, Colloids Surf. 65 (2) (1992) 131–137.
[43] P.A. Kralchevsky, N.D. Denkov, K.D. Danov, Particles with an undulated contact line at a fluid interface: Interaction between capillary quadrupoles and rheology of particulate monolayers, Langmuir 17 (24) (2001) 7694–7705.
[44] J. Fournier, P. Galatola, Anisotropic capillary interactions and jamming of colloidal particles trapped at a liquid–fluid interface, Phys. Rev. E 65 (2002) 031601.
[45] K.D. Danov, P.A. Kralchevsky, B.N. Naydenov, G. Brenn, Interactions between particles with an undulated contact line at a fluid interface: Capillary multipoles of arbitrary order, J. Colloid Interface Sci. 287 (1) (2005) 121–134.
[46] K.D. Danov, P.A. Kralchevsky, Capillary forces between particles at a liquid interface: General theoretical approach and interactions between capillary multipoles, Adv. Colloid Interface Sci. 154 (1–2) (2010) 91–103.
[47] P.V. Petkov, K.D. Danov, P.A. Kralchevsky, Monolayers of charged particles in a langmuir trough: could particle aggregation increase the surface pressure?, Journal of colloid and interface science 462 (2016) 223–234
[48] K.D. Danov, P.A. Kralchevsky, Electric forces induced by a charged colloid particle attached to the water–nonpolar fluid interface, J. Colloid Interface Sci. 298 (1) (2006) 213–231.
[49] M. Oettel, S. Dietrich, Colloidal interactions at fluid interfaces, Langmuir 24 (2008) 1425–1441.
[50] K.D. Danov, P.A. Kralchevsky, M.P. Boneva, Electrodipping force acting on solid particles at a fluid interface, Langmuir 20 (15) (2004) 6139–6151.
[51] K.D. Danov, P.A. Kralchevsky, M.P. Boneva, Shape of the capillary meniscus around an electrically charged particle at a fluid interface: comparison of theory and experiment, Langmuir 22 (6) (2006) 2653–2667.
[52] M.P. Boneva, N.C. Christov, K.D. Danov, P.A. Kralchevsky, Effect of electric-field-induced capillary attraction on the motion of particles at an oil–water interface, PCCP 9 (48) (2007) 6371–6384.
[53] M.P. Boneva, K.D. Danov, N.C. Christov, P.A. Kralchevsky, Attraction between particles at a liquid interface due to the interplay of gravity- and electric-field-induced interfacial deformations, Langmuir 25 (16) (2009) 9129–9139.
[54] K. Danov, P.A. Kralchevsky, Interaction between like-charged particles at a liquid surface: Electrostatic repulsion vs. capillary attraction, J. Colloid Interface Sci. 345 (2010) 505–514.
[55] L. Foret, A. Wurger, Electric-field induced capillary interaction of charged particles at a polar interface, Phys. Rev. Lett. 92 (2004) 058302.
[56] A. Wurger, L. Foret, Capillary attraction of colloid particles at an aqueous interface, J. Phys. Chem. B 109 (2005) 16435–16438.
[57] N. Vandewalle, L. Clermont, D. Terwagne, S. Dorbolo, E. Mersch, G. Lumay, Symmetry breaking in a few-body system with magnetocapillary interactions, Phys. Rev. E 85 (4) (2012) 041402.
[58] K. Danov, R. Aust, F. Durst, U. Lange, On the slow motion of an interfacial viscous droplet in a thin liquid layer, Chem. Eng. Sci. 50 (1995) 2943–2956.
[59] K.D. Danov, R. Dimova, B. Pouligny, Viscous drag of a solid sphere straddling a spherical or flat surface, Physics of Fluids (1994-present) 12 (11) (2000) 2711–2722.
[60] T.M. Fischer, P. Dhar, P. Heinig, The viscous drag of spheres and filaments moving in membranes or monolayers, J. Fluid Mech. 558 (2006) 451–475.
[61] C. Pozrikidis, Particle motion near and inside an interface, J. Fluid Mech. 575 (2007) 333–357.
[62] A. Dani, G. Keiser, M. Yeganeh, C. Maldarelli, Hydrodynamics of particles at an oil-water interface, Langmuir 31 (49) (2015) 13290–13302.
[63] A. Dörr, S. Hardt, H. Masoud, H.A. Stone, Drag and diffusion coefficients of a spherical particle attached to a fluid–fluid interface, J. Fluid Mech. 790 (2016) 607–618.

[64] S. Das, J. Koplik, R. Farinato, D. Nagaraj, C. Maldarelli, P. Somasundaran, The translational and rotational dynamics of a colloid moving along the air-liquid interface of a thin film, Scientific reports 8 (1) (2018) 8910.

[65] S. Das, J. Koplik, P. Somasundaran, C. Maldarelli, Pairwise hydrodynamic interactions of spherical colloids at a gas-liquid interface, Journal of Fluid Mechanics 915.

[66] A. Read, S.K. Kandy, I.B. Liu, R. Radhakrishnan, K.J. Stebe, Dimerization and structure formation of colloids via capillarity at curved fluid interfaces, Soft matter 16 (25) (2020) 5861–5870.

[67] A. Snezhko, I. Aranson, W.-K. Kwok, Surface wave assisted self-assembly of multidomain magnetic structures, Physical review letters 96 (7) (2006) 078701.

[68] K. Zahn, G. Maret, Two-dimensional colloidal structures responsive to external fields, Current opinion in colloid & interface science 4 (1) (1999) 60–65.

[69] M. Kollmann, R. Hund, B. Rinn, G. Nägele, K. Zahn, H. König, G. Maret, R. Klein, J.K. Dhont, Structure and tracer-diffusion in quasi–two-dimensional and strongly asymmetric magnetic colloidal mixtures, EPL (Europhysics Letters) 58 (6) (2002) 919.

[70] S.G. Booth, R.A. Dryfe, Assembly of nanoscale objects at the liquid/liquid interface, The Journal of Physical Chemistry C 119 (41) (2015) 23295–23309.

[71] N.D. Denkov, O. Velev, P.A. Kralchevsky, I.B. Ivanov, H. Yoshimura, K. Nagayama, Two dimensional crystallization, Nature (London) 361 (1993) 26.

[72] B.-J. Lin, L.-J. Chen, Phase transitions in two-dimensional colloidal particles at oil/water interfaces, The Journal of chemical physics 126 (3) (2007) 034706.

[73] N.G. Chisholm, K.J. Stebe, Driven and active colloids at fluid interfaces, Journal of Fluid Mechanics 914.

[74] A.J. Mendoza, E. Guzmán, F. Martínez-Pedrero, H. Ritacco, R.G. Rubio, F. Ortega, V.M. Starov, R. Miller, Particle laden fluid interfaces: dynamics and interfacial rheology, Adv. Colloid Interface Sci. 206 (2014) 303–319.

[75] M. Molaei, N.G. Chisholm, J. Deng, J.C. Crocker, K.J. Stebe, Interfacial flow around brownian colloids, Phys. Rev. Lett. 126 (22) (2021) 228003.

[76] M. Pourali, M. Kröger, J. Vermant, P.D. Anderson, N.O. Jaensson, Drag on a spherical particle at the air–liquid interface: Interplay between compressibility, marangoni flow, and surface viscosities, Phys. Fluids 33 (6) (2021) 062103.

[77] B.J. Park, D. Lee, Particles at fluid-fluid interfaces: From single-particle behavior to hierarchical assembly of materials, MRS Bull. 39 (12) (2014) 1089.

[78] N.D. Vassileva, D. van den Ende, F. Mugele, J. Mellema, Capillary forces between spherical particles floating at a liquid-liquid interface, Langmuir 21 (24) (2005) 11190–11200.

[79] M.-J. Dalbe, D. Cosic, M. Berhanu, A. Kudrolli, Aggregation of frictional particles due to capillary attraction, Phys. Rev. E 83 (5) (2011) 051403.

[80] M.P. Boneva, K.D. Danov, N.C. Christov, P.A. Kralchevsky, Attraction between particles at a liquid interface due to the interplay of gravity-and electric-field-induced interfacial deformations, Langmuir 25 (16) (2009) 9129–9139.

[81] A. Dörr, S. Hardt, Driven particles at fluid interfaces acting as capillary dipoles, J. Fluid Mech. 770 (2015) 5–26.

[82] A. Dani, G. Keiser, M. Yeganeh, C. Maldarelli, Hydrodynamics of particles at an oil–water interface, Langmuir 31 (49) (2015) 13290–13302.

[83] N. Laal Dehghani, R. Khare, G.F. Christopher, 2d stokesian approach to modeling flow induced deformation of particle laden interfaces, Langmuir 34 (3) (2017) 904–916.

[84] S. Barman, G.F. Christopher, Role of capillarity and microstructure on interfacial viscoelasticity of particle laden interfaces, J. Rheol. 60 (1) (2016) 35–45.

[85] N. Laal-Dehghani, G.F. Christopher, 2d stokesian simulation of particle aggregation at quiescent air/oil-water interfaces, Journal of colloid and interface science 553 (2019) 259–268.

[86] S.E. Rahman, N. Laal-Dehghani, G.F. Christopher, Interfacial viscoelasticity of self-assembled hydrophobic/hydrophilic particles at an air/water interface, Langmuir 35 (40) (2019) 13116–13125.

[87] S.E. Rahman, N. Laal-Dehghani, S. Barman, G.F. Christopher, Modifying interfacial interparticle forces to alter microstructure and viscoelasticity of densely packed particle laden interfaces, Journal of colloid and interface science 536 (2019) 30–41.

[88] A. Vidal, L. Botto, Slip flow past a gas–liquid interface with embedded solid particles, Journal of fluid mechanics 813 (2017) 152–174.

[89] M. De Corato, V. Garbin, Capillary interactions between dynamically forced particles adsorbed at a planar interface and on a bubble, Journal of fluid mechanics 847 (2018) 71–92.

[90] A. Huerre, M. De Corato, V. Garbin, Dynamic capillary assembly of colloids at interfaces with 10,000 g accelerations, Nat. Commun. 9 (1) (2018) 3620.

[91] H. Nishikawa, S. Maenosono, Y. Yamaguchi, T. Okubo, Self-assembling process of colloidal particles into two-dimensional arrays induced by capillary immersion force: A simulation study with discrete element method, J. Nanopart. Res. 5 (1–2) (2003) 103–110.

[92] M. Fujita, H. Nishikawa, T. Okubo, Y. Yamaguchi, Multiscale simulation of two-dimensional self-organization of nanoparticles in liquid film, Japanese journal of applied physics 43 (7R) (2004) 4434.

[93] H. Nishikawa, M. Fujita, S. Maenosono, Y. Yamaguchi, T. Okudo, Effects of frictional force on the formation of colloidal particle monolayer during drying–study using discrete element method–[translated], KONA Powder and Particle Journal 24 (2006) 192–202.

[94] P.C. Millett, Y.U. Wang, Diffuse-interface field approach to modeling arbitrarily-shaped particles at fluid–fluid interfaces, Journal of colloid and interface science 353 (1) (2011) 46–51.

[95] A. Uzi, Y. Ostrovski, A. Levy, Modeling and simulation of particles in gas–liquid interface, Adv. Powder Technol. 27 (1) (2016) 112–123.

[96] N. Vandewalle, N. Obara, G. Lumay, Mesoscale structures from magnetocapillary self-assembly, Eur. Phys. J. E 36 (10) (2013) 1–6.

[97] G. Lumay, N. Obara, F. Weyer, N. Vandewalle, Self-assembled magnetocapillary swimmers, Soft Matter 9 (8) (2013) 2420–2425.

[98] A. Darras, F. Mignolet, N. Vandewalle, G. Lumay, Remote-controlled deposit of superparamagnetic colloidal droplets, Phys. Rev. E 98 (6) (2018) 062608.

[99] K. Zahn, J.M. Méndez-Alcaraz, G. Maret, Hydrodynamic interactions may enhance the self-diffusion of colloidal particles, Physical review letters 79 (1) (1997) 175.

[100] B. Rinn, K. Zahn, P. Maass, G. Maret, Influence of hydrodynamic interactions on the dynamics of long-range interacting colloidal particles, EPL (Europhysics Letters) 46 (4) (1999) 537.

[101] H. Löwen, R. Messina, N. Hoffmann, C.N. Likos, C. Eisenmann, P. Keim, U. Gasser, G. Maret, R. Goldberg, T. Palberg, Colloidal layers in magnetic fields and under shear flow, J. Phys.: Condens. Matter 17 (45) (2005) S3379.

[102] E. Hilou, D. Du, S. Kuei, S.L. Biswal, Interfacial energetics of two-dimensional colloidal clusters generated with a tunable anharmonic interaction potential, Physical Review Materials 2 (2) (2018) 025602.

[103] D. Du, E. Hilou, S.L. Biswal, Reconfigurable paramagnetic microswimmers: Brownian motion affects non-reciprocal actuation, Soft matter 14 (18) (2018) 3463–3470.

[104] J. Rotne, S. Prager, Variational treatment of hydrodynamic interaction in polymers, J. Chem. Phys. 50 (11) (1969) 4831–4837.

[105] R. Pesché, G. Nägele, Dynamical properties of wall-confined colloids, EPL (Europhysics Letters) 51 (5) (2000) 584.

[106] R. Pesché, G. Nägele, Stokesian dynamics study of quasi-two-dimensional suspensions confined between two parallel walls, Physical Review E 62 (4) (2000) 5432.

[107] Q. Xie, G.B. Davies, J. Harting, Controlled capillary assembly of magnetic janus particles at fluid–fluid interfaces, Soft Matter 12 (31) (2016) 6566–6574.

[108] Q. Xie, G.B. Davies, J. Harting, Direct assembly of magnetic janus particles at a droplet interface, ACS nano 11 (11) (2017) 11232–11239.

[109] K.D. Danov, P.A. Kralchevsky, Electric forces induced by a charged colloid particle attached to the water–nonpolar fluid interface, Journal of colloid and interface science 298 (1) (2006) 213–231.

[110] H.J. Keh, S.B. Chen, Particle interactions in electrophoresis: I. motion of two spheres along their line of centers, Journal of colloid and interface science 130 (2) (1989) 542–555.

[111] M. O'Neill, Exact solutions of the equations of slow viscous flow generated by the asymmetrical motion of two equal spheres, Appl. Sci. Res. 21 (1) (1969) 452–466.

[112] A. Goldman, R. Cox, H. Brenner, The slow motion of two identical arbitrarily oriented spheres through a viscous fluid, Chem. Eng. Sci. 21 (12) (1966) 1151–1170.

[113] A.J. Banchio, J.F. Brady, Accelerated stokesian dynamics: Brownian motion, The Journal of chemical physics 118 (22) (2003) 10323–10332.

[114] P.A. Kralchevsky, K.D. Danov, P.V. Petkov, Soft electrostatic repulsion in particle monolayers at liquid interfaces: surface pressure and effect of aggregation, Philosophical Transactions of the Royal Society A: Mathematical, Physical and Engineering Sciences 374 (2072) (2016) 20150130.

[115] I. Muntz, F. Waggett, M. Hunter, A.B. Schofield, P. Bartlett, D. Marenduzzo, J.H. Thijssen, Interaction between nearly hard colloidal spheres at an oil-water interface, Physical Review Research 2 (2) (2020) 023388.

[116] J. Deng, M. Molaei, N.G. Chisholm, K.J. Stebe, Interfacial flow around a pusher bacterium, arXiv preprint arXiv:2204.02300.

[117] Z. Qiu, A. Malfliet, B. Blanpain, M. Guo, Capillary interaction between micron-sized ce2o3 inclusions at the ar gas/liquid steel interface, Metall. Mater. Trans. B (2022) 1–17.

[118] C. Albert, N. Huang, N. Tsapis, S. Geiger, V. Rosilio, G. Mekhloufi, D. Chapron, B. Robin, M. Beladjine, V. Nicolas, et al., Bare and sterically stabilized plga nanoparticles for the stabilization of pickering emulsions, Langmuir 34 (46) (2018) 13935–13945.

[119] V. Lotiti, T. Zambelli, Approaches to self-assembly of colloidal monolayers: A guide for nanotechnologists, Adv. Colloid Interface Sci. 246 (2017) 217–274.

[120] F. Martínez-Pedrero, P. Tierno, Advances in colloidal manipulation and transport via hydrodynamic interactions, Journal of colloid and interface science 519 (2018) 296–311.

[121] D. Pérez-Juárez, R. Sánchez, P. Díaz-Leyva, A. Kozina, Equilibrium clustering of colloidal particles at an oil/water interface due to competing long-range interactions, J. Colloid Interface Sci. 571 (2020) 232–238.

[122] V. Carrasco-Fadanelli, R. Castillo, Measurement of the force between uncharged colloidal particles trapped at a flat air/water interface, Soft Matter 15 (29) (2019) 5815–5818.

[123] L.H. Eun, C.K. Hwan, M. Xia, K.D. Woo, P.B. Jun, Interactions between polystyrene particles with diameters of several tens to hundreds of micrometers at the oil–water interface, J. Colloid Interface Sci. 560 (2020) 838–848.

[124] R. Aveyard, B. Binks, J.H. Clint, P.D.I. Fletcher, T. Horozov, B. Neumann, V. Paunov, J. Annesley, S. Botchway, D. Ness, A. Parker, A. Ward, A.N. Burgess, Measurement of long-range repulsive forces between charged particles at an oil-water interface, Phys. Rev. Lett. 88 (2002) 246102.

[125] C. Pozrikidis, Particle motion near and inside an interface, J. Fluid Mech. 575 (2007) 333–357.
[126] J.-C. Loudet, M. Qiu, J. Hemauer, J. Feng, Drag force on a particle straddling a fluid interface: Influence of interfacial deformations, Eur. Phys. J. E 43 (2) (2020) 1–13.
[127] N. Ben'MBarek, A. Aschi, C. Blanc, M. Nobili, Microspheres viscous drag at a deformed fluid interface: particle's weight and electrical charges effects, Eur. Phys. J. E 44 (2) (2021) 1–6.